%% file: main.tex
\documentclass[conference]{IEEEtran}
\IEEEoverridecommandlockouts
\usepackage{easyReview}
\usepackage{cite}
\usepackage{amsmath,amssymb,amsfonts}
\usepackage{algorithmic}
\usepackage{graphicx}
\usepackage{textcomp}
\usepackage{xcolor}
\usepackage{array}
\usepackage{subfloat}
\usepackage{setspace}
\usepackage{makecell} 
\usepackage{hyperref}
\def\BibTeX{{\rm B\kern-.05em{\sc i\kern-.025em b}\kern-.08em
    T\kern-.1667em\lower.7ex\hbox{E}\kern-.125emX}}

\begin{document}

\title{\LARGE Scaling Down to Scale Up: A Cost-Benefit Analysis of Replacing OpenAI’s LLM with Open Source SLMs in Production\\
}

\iftrue
\author{\IEEEauthorblockN{Chandra Irugalbandara}
\IEEEauthorblockA{\textit{Jaseci Labs} \\
chandra.irugalbandara@jaseci.org}
\and
\IEEEauthorblockN{Ashish Mahendra}
\IEEEauthorblockA{\textit{Jaseci Labs} \\
ashish.mahendra@jaseci.org}
\and
\IEEEauthorblockN{Roland Daynauth}
\IEEEauthorblockA{\textit{University of Michigan} \\
daynauth@umich.edu}
\and
\IEEEauthorblockN{Tharuka Kasthuri Arachchige}
\IEEEauthorblockA{\textit{Jaseci Labs} \\
tharuka@jaseci.org}
\and
\IEEEauthorblockN{Jayanaka Dantanarayana}
\IEEEauthorblockA{\textit{Jaseci Labs} \\
jayanaka.dantanarayana@jaseci.org}
\and
\IEEEauthorblockN{Krisztian Flautner}
\IEEEauthorblockA{\textit{University of Michigan} \\
manowar@umich.edu}
\and
\IEEEauthorblockN{Lingjia Tang}
\IEEEauthorblockA{\textit{University of Michigan} \\
\textit{Jaseci Labs}\\
lingjia@umich.edu}
\and
\IEEEauthorblockN{Yiping Kang}
\IEEEauthorblockA{\textit{University of Michigan} \\
\textit{Jaseci Labs}\\
ypkang@umich.edu}
\and
\IEEEauthorblockN{Jason Mars}
\IEEEauthorblockA{\textit{University of Michigan} \\
\textit{Jaseci Labs}\\
profmars@umich.edu}
}
\fi

\maketitle
\thispagestyle{plain}
\pagestyle{plain}

\begin{abstract}
Many companies use large language models (LLMs) offered as a service, like OpenAI's GPT-4, to create AI-enabled product experiences. Along with the benefits of ease-of-use and shortened time-to-solution, this reliance on proprietary services has downsides in model control, performance reliability, uptime predictability, and cost. At the same time, a flurry of open-source small language models (SLMs) has been made available for commercial use. However, their readiness to replace existing capabilities remains unclear, and a systematic approach to holistically evaluate these SLMs is not readily available. This paper presents a systematic evaluation methodology and a characterization of modern open-source SLMs and their trade-offs when replacing proprietary LLMs for a real-world product feature. We have designed SLaM, an open-source automated analysis tool that enables the quantitative and qualitative testing of product features utilizing arbitrary SLMs. Using SLaM, we examine the quality and performance characteristics of modern SLMs relative to an existing customer-facing implementation using the OpenAI GPT-4 API. Across 9 SLMs and their 29 variants, we observe that SLMs provide competitive results, significant performance consistency improvements, and a cost reduction of 5$\times$$\sim$29$\times$ when compared to GPT-4.
\end{abstract}


\input{intro}
\input{background}
\input{problem}
\input{tool}

\input{eval}
\input{reference}

\section{Conclusion}
We show that while LLMs, like GPT-4, have excellent breadth of capabilities, their power can be approximated with smaller, faster, and cheaper models for domain-specific tasks. We evaluate the trade-offs on the part of a commercial application that generates a daily "pep talk" to its user based on their past behavior and future tasks. While GPT-4 achieves the highest accuracy across a broad range of models, as judged by a human panel, most small language models (SLMs) come close to its response quality while incurring 1/5 to 1/29 of the cost. The SLMs also benefit from more predictable latency performance. Our evaluation is facilitated by an automated methodology and open-source toolset called SLaM, which facilitates the acquisition and hosting of the SLMs, and conducts model response quality evaluation and provides performance and cost analysis of the SLMs.

\bibliographystyle{IEEEtran}
\bibliography{references.bib}

\end{document}

%% file: intro.tex
\section{Introduction}
Generative AI (GenAI), particularly generative large language models (LLMs), has recently grown in popularity within academic and industry communities.
This surge is largely attributed to OpenAI's launch of ChatGPT~\cite{openai_chatgpt}, and the family of GPT models and their groundbreaking performance and capabilities across a wide range of natural language generation tasks. These tasks range from intelligent chatbots and specialized document creation, to coding assistants and many others. 
Indeed, this emerging class of AI has disrupted the commercial AI landscape as they have become a top strategic priority for many startups and established corporations~\cite{dowling2023chatgpt}.

Generative LLMs take natural language input (i.e., prompts) and generate responses that follow the instructions and information provided in the input prompt.
These models are transformer neural networks with billions of parameters and are prohibitively expensive to train and serve.
Consequently, many companies use cloud APIs based models to develop their next-generation GenAI products and features. 

OpenAI's GPT models, especially GPT-4, have emerged as a favored choice for their cutting-edge capabilities and developer-friendly interface~\cite{openai_introducing_chatgpt_whisper_apis}, facilitating rapid prototyping and short time-to-market.
However, reliance on proprietary cloud APIs to access LLMs' capabilities presents numerous challenges.
Issues such as inconsistent request latency and outages during peak usage times have been noted with OpenAI APIs~\cite{theregister_chatgpt_outage_2023}.
Furthermore, the per-token pricing model can accumulate significant costs over time and at scale, which is particularly challenging for startups.
Additionally, the proprietary nature of cloud APIs limits developers from customizing models with their data to better suit specific use cases through fine-tuning.

Meanwhile, the AI community has seen the emergence of numerous open-source language models of smaller sizes that are available for commercial use~\cite{touvron2023llama, jiang2023mistral, neuralchat, openllms23, mitra2023orca, stablelm, starling2023, vicuna2023, tunstall2023zephyr}. 
Several quantization techniques have recently been published and applied to make these models even smaller~\cite{shen2023efficient}\cite{chang2023effective}. 
We categorize these models as Small Language Models (SLMs).
These SLMs have shown promising generative performance, comparable to larger LLMs in certain benchmarks~\cite{jiang2023mistral}\cite{wang2023openchat}\cite{mukherjee2023orca}.
However, the readiness of these SLMs to replace proprietary LLMs (e.g. OpenAI GPT-4) in production settings remains unclear, particularly regarding response quality, performance, and cost-effectiveness. 
Additionally, a methodology and tooling for evaluating SLMs for a particular product feature is not readily available.

This work presents SLaM, a systematic methodology and tooling for evaluating open-source SLMs compared to proprietary LLM APIs. SLaM (1) automates the acquisition and hosting of SLMs into a local or cloud environment, (2) provides human-in-the-loop tooling for evaluation of model response quality, and (3) includes performance and cost analysis. Using SLaM, we characterize modern SLMs and their quantized variants through a case study of replacing a production application's existing OpenAI GPT-4-based feature. We examine four questions relevant to the viability of replacing GPT-4 with self-hosted SLMs:
\begin{enumerate}
    \item Is the quality of response from modern SLMs good enough for users?
    \item How well can AI-assisted tooling automate the process of evaluating SLMs?
    \item What are the latency implications of self-hosted SLMs in a utility-based cloud environment such as AWS?
    \item What are the cost trade-offs of self-hosted SLMs compared to using third-party cloud APIs?
\end{enumerate}

Utilizing the SLaM evaluation methodology and its accompanying toolset, we analyze a suite of 9 distinct SLMs and their quantized versions, totaling 29 different models.
We evaluate the trade-offs of applying these models to a real user-facing feature in myca.ai~\cite{mycaai}, an AI-powered productivity tool currently in-production.
The feature is a~\textit{daily pep talk} feature, where it leverages the user's tasks list to deliver personalized and intelligent encouragement and advice on a daily basis.
The current version of this feature is built using OpenAI GPT-4 and in production.
In this case study, we show that open-source SLMs can effectively replace OpenAI's models in a production environment.
Specifically, we find that SLMs can generate responses with similar levels of quality, provide a comparable yet more reliable latency performance, and reduce costs by up to 29$\times$.
This paper makes the following contributions:
\begin{itemize}
    \item We introduce a systematic methodology for evaluating open-source SLMs for AI  feature implementation and comparing them to proprietary LLMs.
    \item We design and develop the SLaM tool, which supports both human-in-the-loop and AI-assisted automated evaluation processes. SLaM is open-source and available at~\url{https://github.com/Jaseci-Labs/slam}.
    \item We present a characterization of modern quantized SLMs through a case study of replacing GPT-4 in a real-world application. We find SLMs are indeed viable alternatives to proprietary LLMs and have advantages such as more reliable latency and lower costs.
\end{itemize}

%% file: background.tex
\begin{figure}[t!]
\centering
\includegraphics[width=0.95\columnwidth]{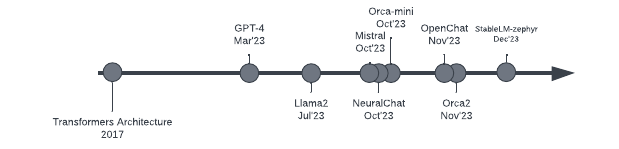}
\caption {Brief history of evolution of language models and recent surge in open-source SLMs.}
\label{fig:history_llms}
\end{figure}

\section{Background: The Recent Evolution of LLMs}

\subsection{Large Language Models}
LLMs are language models that are pre-trained on a large corpus of data to process, comprehend and generate natural language. 
Because of their outstanding performance in many application domains, LLMs have gained tremendous popularity in both academia and industry~\cite{chang2023survey}.
LLMs can be applied to various tasks, such as text generation and reasoning, as well as to feature implementation in real-world applications, such as virtual assistants, chatbots, and language translation systems~\cite{Ollivier2023}.

In contrast to the conventional approach for NLP tasks, which often involves fine-tuning models through supervised learning on task-specific datasets, LLMs can effectively perform a wide range of tasks based on instructions in the input (prompts) without additional training.
By providing LLMs with instructions and/or examples of the desired outcome, they can perform tasks for which they have not been explicitly trained for previously.

Vaswani et al.~\cite{vaswani2023attention} first introduced in 2017 the transformer architecture and self-attention mechanism.
This key architecture innovation led to a series of transformer-based language models that achieve state-of-the-art performance across many natural language tasks.
GPT-2~\cite{Radford2019LanguageMA}, introduced in 2019, is a large transformer-based model with 1.5 billion parameters.
Megatron-LM (2019)~\cite{shoeybi2020megatronlm} surpasses GPT-2 with 8.3 billion parameters. 
This extensive size enables Megatron-LM to capture and generate more intricate linguistic patterns.
Subsequently, OpenAI introduced GPT-3 in July 2020, GPT-3.5 in March 2022 and GPT-4, their latest LLM, in March 2023.
The family of GPT models were considered the most advanced language models upon release.
They power many of OpenAI’s popular APIs and applications, including the widely used ChatGPT, which has attracted a significant amount of main-stream attention to LLM and AI research in general~\cite{OpenAI_GPT4_2023}.


\subsection{Impact of LLMs}
LLMs are revolutionizing the tech industry by altering how we work and interact with information.
LLMs' human-like capabilities of understanding and generating natural language text have been leveraged to assist and power many application use cases, including virtual assistants, language translation, content writing, and scientific research.
Significant increases in efficiency and the establishment of new job categories are made possible with the adoption of LLMs.
Over 2 million developers have adopted LLMs for their applications.~\cite{openAIdevday}.



\subsection{Proprietary LLMs and OpenAI APIs}
OpenAI APIs have quickly emerged as the preferred cloud option for LLM inference.
While OpenAI’s APIs provide convenient access to powerful language models such as GPT-4, reliance on cloud APIs and proprietary models presents several challenges for developers, including lack of model control, unreliable uptime, and potentially significant cost. 
Furthermore, cloud APIs also limit the developer's ability to fine-tune the models with custom data to tailor to specific tasks 

\subsection{Open-source Small Language Models}
There has been a surge of open-source language models released by the research community and industry companies.
Figure~\ref{fig:history_llms} highlights the timeline of several key open-source LLMs released since the launch of ChatGPT and release of GPT-4.
These models are generally smaller in size, making them more feasible to self-hosting.
In addition, several quantization techniques have been proposed to make these models even smaller without significant sacrifice to accuracy~\cite{liu2023llmqat, liu2023qllm}.
The quantization process entails mapping the model's weights from data types with higher precision (16bit) to ones with lesser precision (4bit, 3bit, 2bit), therefore effectively compressed the model and reduces their memory requirements.

\subsection{Developing with LLMs}
The open-source community has created new infrastructure and tools to make developing with LLM an easy process, further accelerating the adoption of LLMs into modern applications and products.
LangChain~\cite{langchain-ai} is a framework that assists developers with creating multi-step reasoning pipelines using language models.
LlamaIndex~\cite{llamaindex} is a framework that focuses on the integration between LLMs with bespoke data sources. It facilitates the ingestion, indexing and preparation of different data sources and structures. 

%% file: problem.tex
\section{Problem: Realizing the "Daily Pep Talk" Feature}

\begin{figure}[t] 

  \begin{minipage}[b]{0.49\linewidth}
    \centering
    \includegraphics[width=\linewidth]{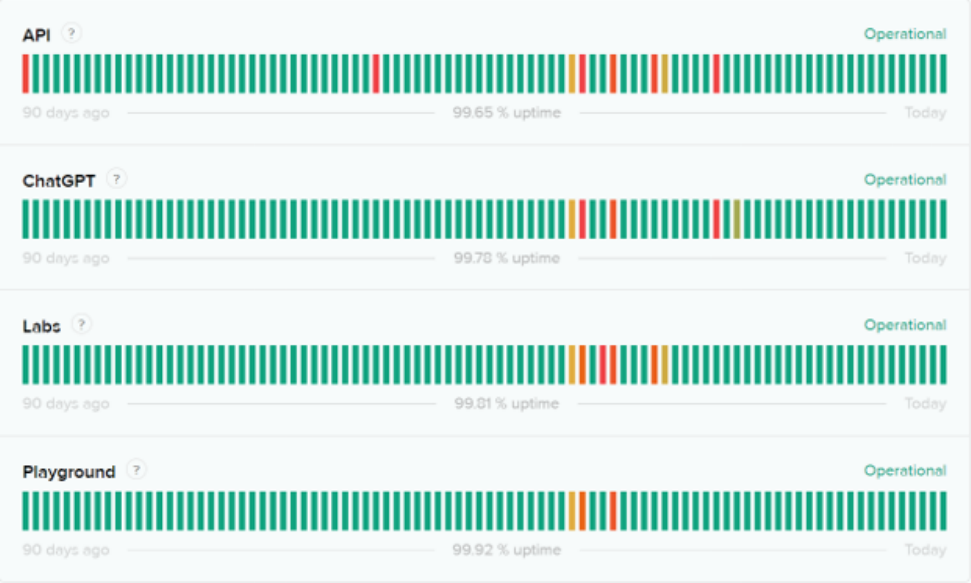}
    \caption{OpenAI APIs status, captured on  12/14/23}
    \label{fig:openai_outages}
  \end{minipage}
  \hfill 
  \begin{minipage}[b]{0.49\linewidth}
    \centering
    \includegraphics[width=\linewidth]{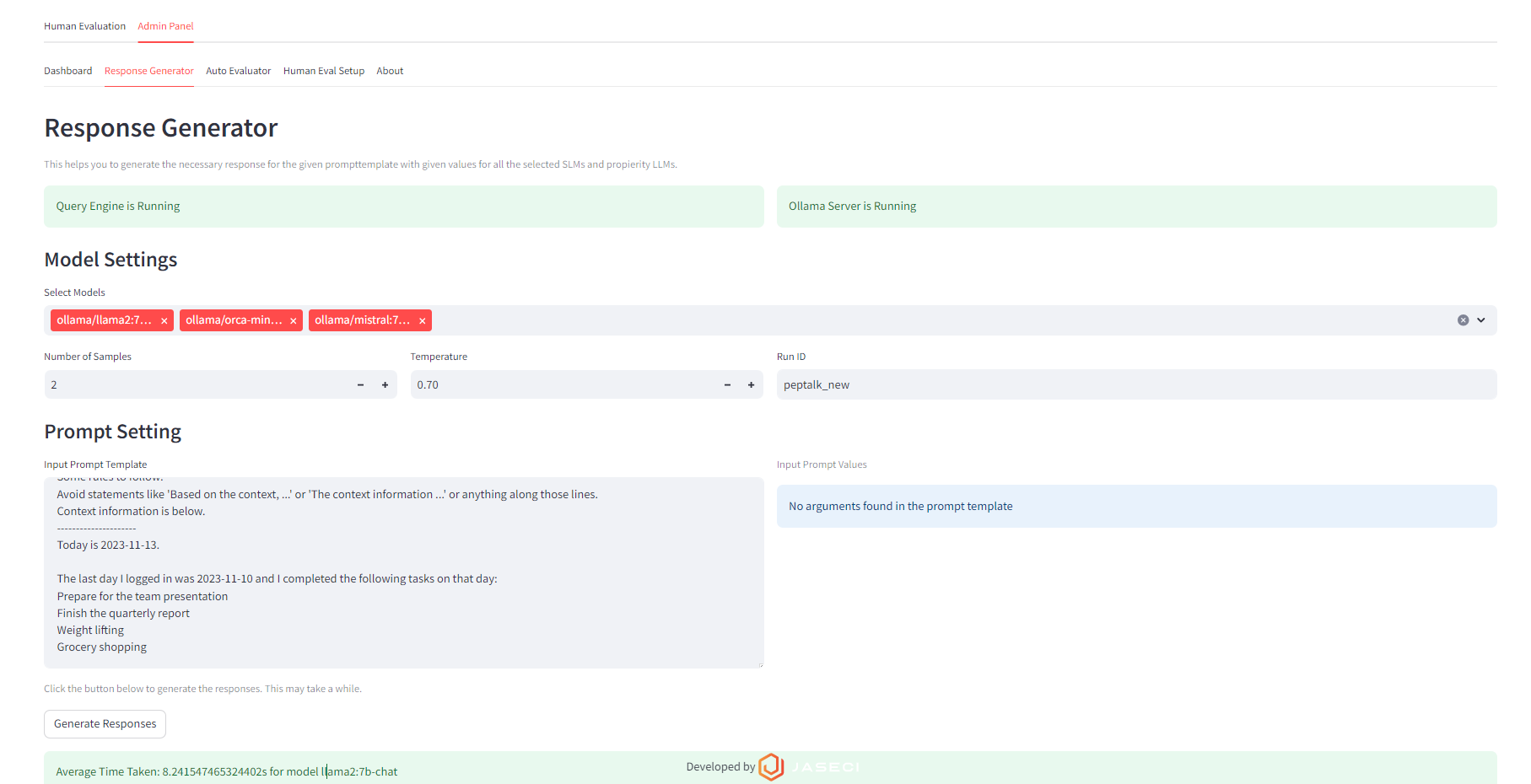}
    \caption{SLaM Tool UI Interface}
    \label{fig:SLAM_analyzer}
  \end{minipage}
\end{figure}

\label{sec:problem}
This paper investigates the feasibility of replacing an LLM (via OpenAI API) with SLMs in a production use case. In particular, we are interested in the response quality of SLMs compared to LLMs. In addition, we aim to understand whether SLMs can address and mitigate key limitations of proprietary LLM APIs, specifically cost and latency.

We conduct a case study of a real AI feature in a production application. We first describe the details of the application and the feature. We then present the current key limitations of using OpenAI APIs in production and the research questions we aim to answer when comparing SLMs with LLMs. 

\subsection{Product Feature Case Study}
The application in this case study is myca.ai~\cite{mycaai}, a personal task-management and productivity application that has been in production for over a year. 
In myca.ai, users create and manage their plan and tasks across all aspects of their life, such as work, personal health and finances, to stay organized, focused and productive.
In addition, users can set their longer-term goals and daily habits.
myca.ai records for the user what are accomplished daily, their progress towards goals and habits patterns, etc.

myca.ai has a number of AI features that assist its users in improving their productivity.
One key AI feature is the ``Daily Pep Talk".  
At the beginning of every day, myca provides an encouraging message to the user based on what they accomplished the day before, what they plan to do today, and progress towards their goals.

To generate this message, user's activity from the previous day, along with their plan for the coming day are combined with additional context and instructions into a text input to a LLM.
This input is commonly referred to as a "prompt".
The LLM "follows" the instruction in the prompt and generates a response that is then shown to the user.
The process of constructing and tweaking the input prompt to the LLM is commonly called "Prompt Engineering" and is the main method of leveraging a pre-trained LLM to a task-specific use case~\cite{prompt,10.1145/3560815}.
We show the prompt template for the Daily Pep Talk feature below.
The placeholders (marked by brackets) are replaced with user specific information before sending to LLM for inference.

\vspace{6pt}
{
\scriptsize
\fontfamily{cmbr}\selectfont
\begin{spacing}{0.9}
You are given below a description of the tasks and goals in my todo list. Always answer the query using the provided context information, and not prior knowledge.

Some rules to follow:

1. Avoid statements like 'Based on the context, ...' or 'The context information ...' or anything along those lines.

Context information is below.

---------------------

Today is 2023-11-13.

The last day I logged in was 2023-11-10 and I completed the following tasks on that day:

[LIST OF TASKS COMPLETED]

Here are today's focused tasks. These are tasks that I think are important:

[LIST OF TASKS PLANNED] 

Here are the rituals that are scheduled for today. These are recurring tasks that help me build and maintain good habits or work/life responsibilities that happen regularly:

[LIST OF RECURRING TASKS SCHEDULED] 

I have also set the following goals for myself for this week. These are overarching objectives I want to accomplish in this week:

[LIST OF USER SET GOALS]

---------------------

Given the context information, answer the query.

Query: Imagine you are my personal assistant, generate a short briefing for me at the start of my day. In the briefing, summarize what I completed in the previous day and then give me a preview of the key activities for today. In this briefing, consider my goals for this week and tell me if my focused tasks and rituals are aligned with those goals. Carefully evaluate the associations between the tasks and goals and describe the tasks based on how related you think they are. Note that it is possible that a task is not directly associated with any goals. Reference the specific tasks mentioned in the context and generate this briefing in a single, naturally flowing narrative. Avoid simply listing out tasks one by one. Use a motivating and encouraging tone. Keep your response within 4 sentences.

Answer:
\end{spacing}
}

\vspace{6pt}


\subsection{Challenges with OpenAI APIs}
In the currently deployed version of the Daily Pep Talk feature, we use the GPT-4 model from OpenAI's cloud APIs. While the OpenAI APIs provide state-of-the-art performance and an easy-to-use interface that facilitates quick prototyping, our experience suggests three key limitations when using them in production settings.

\noindent\textbf{Performance and Reliability} Developers and users of OpenAI have reported a large variability in query latency~\cite{OpenAIforum}, ranging from sub 1s to 4 - 5 minutes. In addition to latency variability, OpenAI has reported frequent API outages since its launch.  Figure~\ref{fig:openai_outages} shows a snapshot of the OpenAI cloud platform status as of December 13th, 2023. In the month of November 2023 alone, it experienced four major outages and three partial outages, some of which lasted more than 3 hours.
Given that OpenAI APIs may experience unpredictable latency and stability, these factors pose considerable risks for deployment in a production environment.
    
\noindent \textbf{Token Usage Limits} The token limits in the OpenAI API can be detrimental to production deployments. Currently, there is a limit of 1000 tokens per minute for OpenAI APIs and 2000 tokens per minute in Azure GPT-4 APIs. During our deployment, even with a mandatory wait time of 10 seconds in between requests, we still encountered issues of exceeding token limit and requests were dropped by OpenAPI without processing.
As a result, the actual resulting latency of those requests increased from 300 milliseconds to more than 5 seconds after subsequent retrying. This is not well suited for production usage.

\noindent\textbf{Cost} As of December 2023, OpenAI's API for GPT-4, their most advanced and capable LLM, is at \$0.03 per 1K input tokens and \$0.06 per 1K output tokens. Assuming each request on average consists of 1000 input tokens and 1000 output tokens, one request costs \$0.09.
Based on our experiences, depending on the nature of the application and feature, certain requests can have 2500 to 3000 input tokens which further raise the per-request cost to \$0.15.
An average traffic of 1000 requests per day would lead to the cost of \$2700 per month, or \$32,400 per year. This can easily scale up to \$1,000,000/month when the request load goes above 360,000 requests/day. In other words, if each user only sends five queries/day, spending \$12,000,000/year for openAI API can only support around 71,000 active users. This is a significant expense for companies, especially for start-ups.

\subsection{Replacing OpenAI with SLMs}
To address these production-related limitations, it is imperative for us to investigate the feasibility of replacing LLMs via cloud APIs with Small Language Models (SLMs) that can be self-hosted to reduce cost and increase performance predictability. 
In the rest of the paper, we set out to answer the following questions:

\begin{itemize}
    \item What is the right process for evaluating SLMs to replace LLMs via APIs for a production use case? How much can this process be automated?
    \item Are SLMs capable of generating responses that are of similar quality as OpenAI's GPT-4?
    \item How do SLMs in production perform in terms of latency performance and consistency compared to OpenAI APIs?
    \item What are the cost trade-offs of self-hosted SLMs vs. OpenAI GPT-4?
\end{itemize}

%% file: tool.tex
\begin{figure}[t]
\centering
\includegraphics[width=0.95\columnwidth]{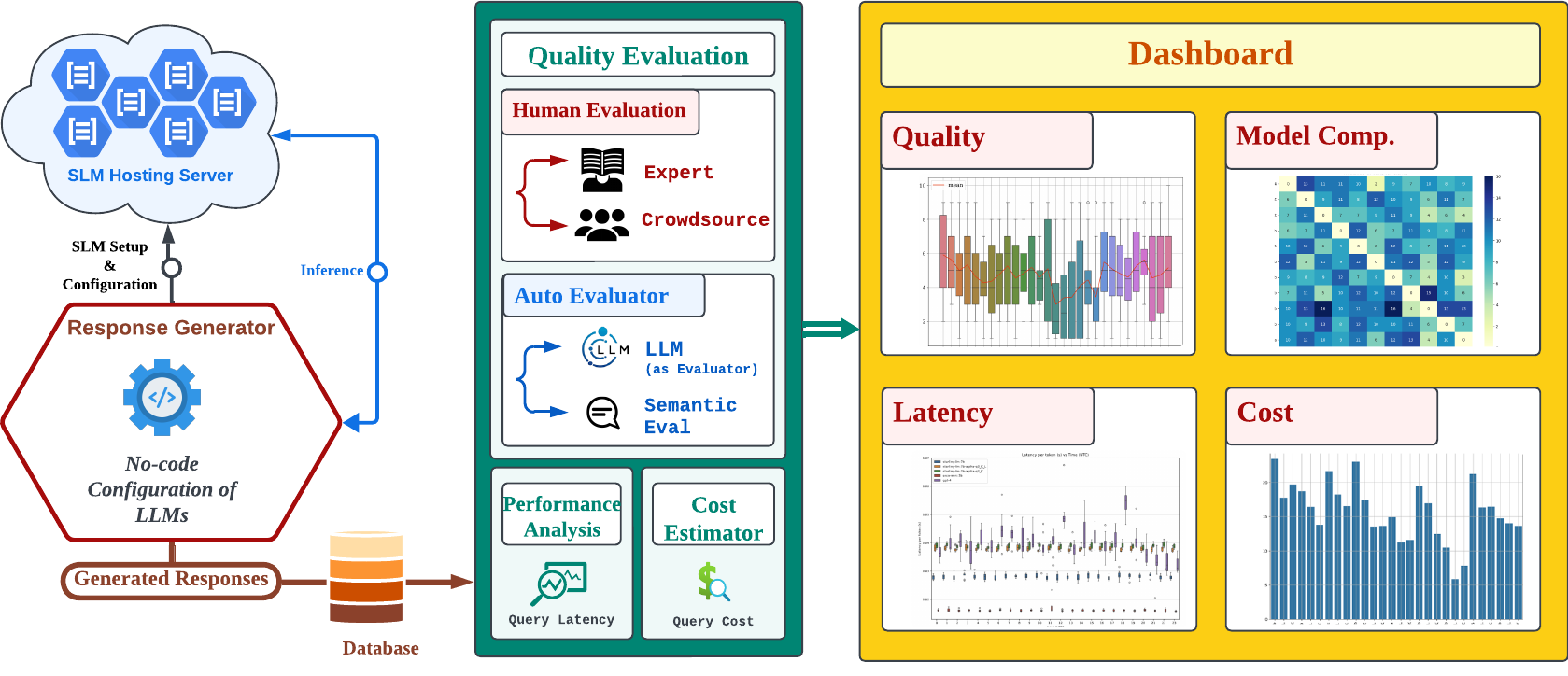}
\caption {Architecture Overview of the SLaM Tool}
\label{fig:slm_bench_tool}
\end{figure}

\begin{figure*}[t] 
  \begin{minipage}[b]{0.35\linewidth}
    \centering
    \includegraphics[width=\linewidth]{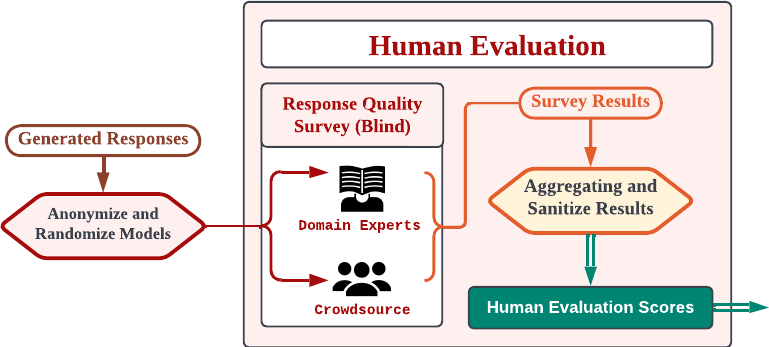}
    \caption{Human Evaluation}
    \label{fig:human_eval_flow}
  \end{minipage}
  \hfill 
  \begin{minipage}[b]{0.26\linewidth}
    \centering
    \includegraphics[width=\linewidth]{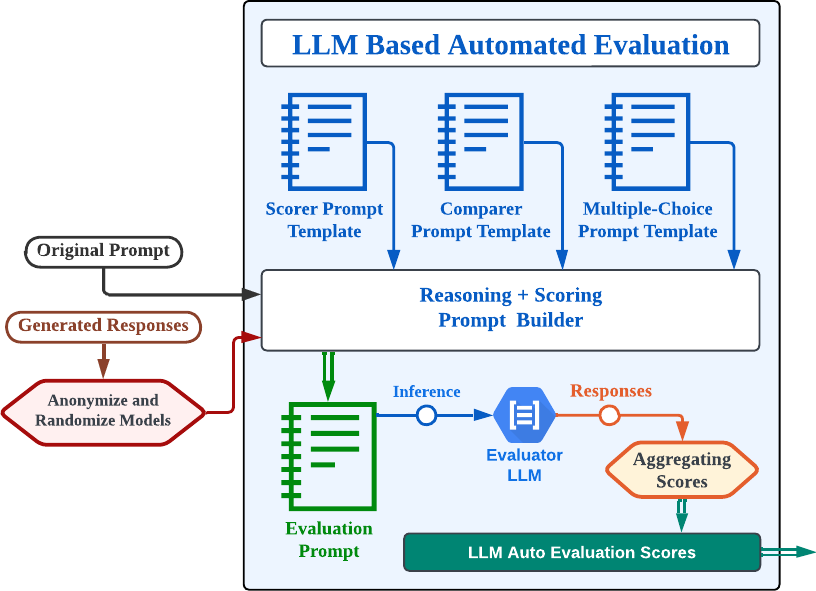}
    \caption{GPT-Based Evaluation}
    \label{fig:gpt4_automated_evaluation_flow}
  \end{minipage}
  \hfill 
  \begin{minipage}[b]{0.34\linewidth}
    \centering
    \includegraphics[width=\linewidth]{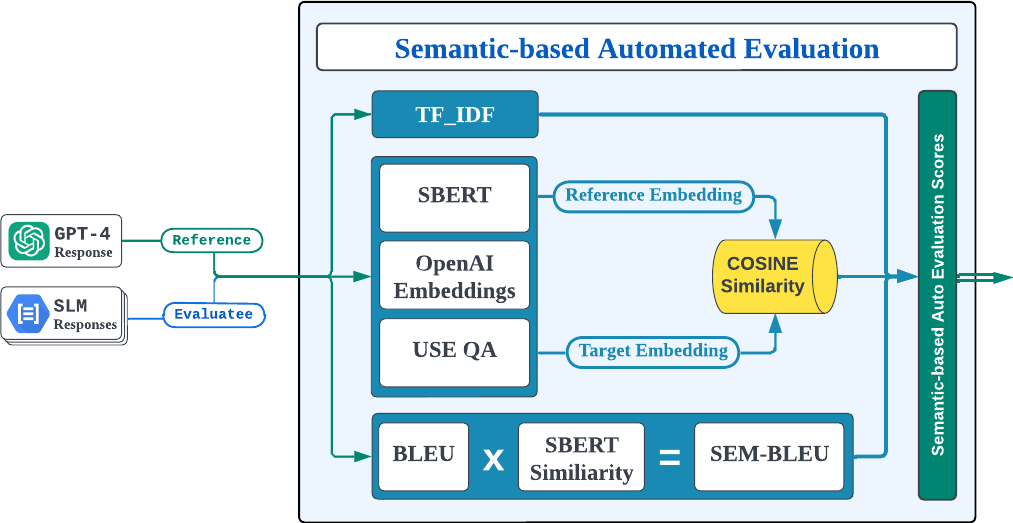}
    \caption{Semantic Similarity Evaluation}
    \label{fig:sem_similarity_fig}
  \end{minipage}
\end{figure*}

\section{SLAM Methodology and Tool}
We present the SLaM toolset, a novel platform to evaluate the performance of SLM vs. LLM across a wide range of metrics (response quality, latency distribution, availability, etc) in production use cases. One of the key contributions of SLaM is to automate and facilitate the evaluation process, including the setup, response generations, measuring, benchmarking, and comparison between SLMs and LLMs, significantly reducing the manual effort required.

\subsection{SLaM Architecture and Components}
SLaM is designed to facilitate a comprehensive analysis of SLMs.
Figure~\ref{fig:slm_bench_tool} illustrates the architecture of SLaM and its key components and Figure~\ref{fig:SLAM_analyzer} shows a screenshot of the user interface of SLaM.
\begin{itemize}
    \item {\textbf{SLM Hosting and Configuration -} SLaM automatically downloads the models of interest from the Hugging Face model repository\cite{HuggingFace} and hosts the models in the AWS cloud. It sets up the model inference for experiments and evaluations.}
    
    \item{\textbf{Human Evaluation -} SLaM facilitates human evaluation of model response quality. For a given input prompt, SLaM collects the response from each candidate model and presents it to the human evaluator through a UI to rate the response quality. The human evaluators are only informed names of the models to ensure a blind tests and unbiased ratings. SLaM can be used for both crowdsource evaluation and custom testing setting up (e.g., experts only).}

    \item{\textbf{Automated Quality Evaluation -} To reduce the effort needed for human evaluation, SLaM provides a set of automated approaches to evaluate response quality. This includes semantic similarity scoring and GPT-4-based scoring. Similarity scoring is designed to provide a quantitative assessment of how closely the responses from an SLM match a reference baseline response while GPT-based scoring conducts rating on the responses similarly to that of a human evaluator. We use a range of similarity metrics, capturing both semantic and syntactic elements of the text.}

    \item{\textbf{Performance and Cost Evaluation -} In addition to model response quality, SLaM provides performance and cost evaluation of the SLMs. We measure average per-request and per-token latency, as well as query latency distribution over 24 hours. We provide cost estimation of self-hosting SLMs and OpenAI APIs.}

    \item {\textbf{Configuration and Dashboard UI -} SLaM features an UI interface for experiment configuration (e.g., problem definition and prompt configuration), as well as a dashboard with various graphs, visualization and analysis of the evaluation outcome.}
\end{itemize}

\subsection{SLaM Response Quality Evaluation Methodology}
\label{subsec:human_eval_methodolodgy}
We describe the SLM response quality evaluation process in SLaM.
\subsubsection{Human Evaluation}
During the evaluation process, each human evaluator is presented with a description of the problem, the original input prompt and the candidate responses generated by the SLMs.
The human evaluator is asked to rate each response based on their judgment of its quality and relevance to the original problem and intention described in the input prompt.
To prevent bias, a blind test is administrated, where the model that generated the response is not disclosed to the evaluator. This ensures that ratings are based solely on the response quality, not on preconceived notions about a model's capabilities or popularity. Responses are presented in randomized order which further helps in reducing evaluation bias. 
Each human evaluator is required to finish the complete set of scorings assigned to them for their response to qualify. This is designed particularly for crowdsource to avoid inattentive and incomplete inputs from the crowd workers. These incomplete responses are removed from the final results as part of the aggregation and sanitation stage of the evaluation pipeline. The final results are visualized in a number of formats in a dashboard. Figure \ref{fig:human_eval_flow} depicts this process.

\subsubsection{GPT-based Evaluation}
\label{subsec:gpt-evaluators}
Human evaluation of response quality, while effective, can be time-consuming.
SLaM includes several automated evaluation techniques that are designed to help reduce the amount of human evaluation required.
We first described three ways that SLaM uses GPT-4 as a tester to evaluate model response quality: Scorer, Comparer and Multi-choice Selector.

\begin{itemize}
    \item \textbf{GPT-Scorer:} GPT-4 is given the prompt and responses from the SLMs and is prompted to rate responses based on its quality and relevance, similar to the instruction given to the human evaluators. The prompt template for the Scorer is as follows: 

    \vspace{2pt}
    {
        \scriptsize
        \fontfamily{cmbr}\selectfont
        \begin{spacing}{0.9}
PROMPT: [prompt]

RESPONSE:[response]

Rate the response on a scale of 0 to 10 (0 being the worst and 10 being the best). Rate the response based on how well it answers the prompt. Reason step by step and then give a score. Only give a number between 0 and 10.
        \end{spacing}
    }
    \vspace{2pt}
    
    \item \textbf{GPT-Comparer:} This methodology instructs GPT-4 to analyze and score responses from SLMs in the context of comparing with a reference baseline response. The reference response is generated by OpenAI API. We prompt the model to also include in the output reasoning behind its decision. We parse the ratings and reasoning from the output and present in the SLaM dashboard alongside visualizations of the ratings. The prompt template for the Comparer is as follows:
    
    \vspace{2pt}
    {
        \scriptsize
        \fontfamily{cmbr}\selectfont
        \begin{spacing}{0.9}
REFERENCE RESPONSE: [reference\_response]

TARGET RESPONSE: [target\_response]

Rate on a scale of 0 to 10 on how close the target response is to the reference response (0 being completely different and 10 being very close in terms of meaning and objective).

Output format

Reason: Reasoning...
 
Score: Rating

    \end{spacing}
    }
    \vspace{2pt}
    
    \item \textbf{GPT-4 Multi-Choice Selector:} GPT-4 is presented with multiple responses from multiple SLMs for a given input prompt. It is then instructed to select the best response in terms of quality, accuracy and relevance. The prompt template for the Multi-choice Selector is as follows:

        \vspace{2pt}
    {
    \scriptsize
    \fontfamily{cmbr}\selectfont
    \begin{spacing}{0.9}
PROMPT:

{prompt}

RESPONSES:

{responses}

What is the best response? Give the choice between 1 and [num\_responses] and provide a short reasoning.

Output format

Reason: Reasoning...

Choice: Choice

    \end{spacing}
    }
    \vspace{2pt}

\end{itemize}
\subsubsection{Semantic Similarity Evaluation}
\label{subsec:similarity-score}
In addition to LLM-based auto-evaluation techniques, we leverage semantic similarity scores to provide a quantitative assessment of how closely the responses of SLMs relate to a reference response. We use a suite of similarity scores including traditional token-based similarity metric and embedding-based similarity measurement.

\begin{itemize}
    \item TF-IDF~\cite{manning2009introduction}: A standard similarity metric used in information retrieval. It assesses the importance of a word to a response in a collection of responses.
    \item SBERT~\cite{DBLP:journals/corr/abs-1908-10084}: We use SBERT to compute the embeddings of each response and calculate cosine similarity of responses.
    \item USE-QA~\cite{cer2018universal}: Similar to SBERT, we calculate the cosine similarity of USE-QA embeddings of the responses.
    \item OpenAI Embedding: Similar to SBERT and USE-QA, embeddings from GPT-4 are used here.
    \item SEM-BLEU: Combines the semantic analysis of SBERT with the syntactic nature of BLEU score~\cite{papineni2002bleu}. The score is the arithmetic mean of the SBERT and BLEU scores. 

\end{itemize}

The goal of SLaM's automated quality evaluation methodology is to provide quick relevant metrics to assess the response quality of SLMs without human ratings. Each of the automated methodologies captures certain aspects of the response quality. We compare and discuss the efficacy of these metrics in Section~\ref{sec:eval}. 




%% file: eval.tex
\section{Experimental Setup}
\subsection{OpenAI API setup}
For OpenAI APIs, we set the temperature parameter as 0.7 and use the default for the rest of the configurations.
The temperature parameter influences the randomness of the generated responses.
We use Langchain~\cite{langchain-ai} to integrate with the OpenAI API.
\subsection{SLMs setup}
\noindent\textbf{Self-hosted AWS instance}
In our experiments, we used the g4.dn.2xlarge AWS EC2 instance type. This instance type has an Intel Xeon processor with 8 virtual CPUs, with 2.5GHz frequency, 32GB of memory, and a single Nvidia T4 GPU with 16GB of dedicated memory.

\noindent\textbf{Frameworks \& libraries}
SLaM leverages Ollama~\cite{ollam} to help facilitate model acquisition, configuration, and inference. 
Ollama operates as a microservice responsible for managing model-related tasks, providing a set of API endpoints for integration with the rest of the SLaM application. 
SLaM constructs the input prompt and sends it to Ollama for SLM inference.

\subsection{SLMs studied in this work}
SLaM is designed to work with any SLMs.
For this case study, we wanted to study SLMs that are representative of state-of-the-art in performance. 
We selected the top models from the Huggingface LLM Leaderboard~\cite{HuggingFaceH4}, including Starling-lm:7b, Mistral-instruct:7b, OpenChat:7b, Zephyr:7b, Stablelm-zephyr:3b, Orca-mini:3b, Vicuna:7b, Orca2:7b, neuralChat:7b, and Llama2-chat:7b.
In addition to the original model, we include the 3-bit and 2-bit quantized versions of these models. 
In total, 29 distinct models are studied, as listed in Table~\ref{tab:slm_studied}.

    \begin{table}[t!]
    \centering
    \scriptsize
    \begin{tabular}{ m{15em}  >{\centering\arraybackslash}m{0.5cm}   >{\centering\arraybackslash}m{0.7cm}   m{2em}  m{2.5em}  m{2.5em}  m{2em} }
    \Xhline{2\arrayrulewidth}
    \scriptsize Model & \scriptsize \# bits & \scriptsize \# Params (B) & \scriptsize Model Size (GB)  & \scriptsize Memory Usage (MB)\\
    \Xhline{2\arrayrulewidth}
    llama2:7b-chat~\cite{touvron2023llama} & 4 & 7 & 3.8 & 6933\\
    
    llama2:7b-chat-q2\_K & 2 & 7 & 2.8 & 5567\\

    llama2:7b-chat-q3\_K{}\_L & 3 & 7 & 3.6 & 6271\\
       \hline
    mistral:7b-instruct~\cite{jiang2023mistral} & 4 & 7 & 4.1 & 4795\\

    mistral:7b-instruct-q2\_K & 2 & 7 & 3.1 & 3897\\

    mistral:7b-instruct-q3\_K\_L & 3 & 7 & 3.8 & 4601\\
       \hline
    neural-chat:7b~\cite{neuralchat} & 4 & 7 & 4.1 & 5183\\

    neural-chat:7b-v3.2-q2\_K & 2 & 7 & 3.1 & 4285\\

    neural-chat:7b-v3.2-q3\_K\_L & 3 & 7 & 3.8 & 4601\\
       \hline
    openchat:7b-v3.5~\cite{wang2023openchat} & 4 & 7 & 4.1 & 5183\\

    openchat:7b-v3.5-q2\_K & 2 & 7 & 3.1 & 4285\\

    openchat:7b-v3.5-q3\_K\_L & 3 & 7 & 3.8 & 4989\\
   \hline
    orca-mini:3b~\cite{mukherjee2023orca} & 4 & 3 & 2 & 3089\\

    orca2:7b~\cite{mitra2023orca} & 4 & 7 & 3.8 & 6527\\

    orca2:7b-q2\_K & 2 & 7 & 2.8 & 5567\\

    orca2:7b-q3\_K\_L & 3 & 7 & 3.6 & 6271\\
   \hline
    stablelm-zephyr:3b~\cite{stablelm} & 4 & 3 & 1.6 & 3443\\

    stablelm-zephyr:3b-q2\_K & 2 & 3 & 1.2 & 3251\\ 

    stablelm-zephyr:3b-q2\_K\_L & 3 & 3 & 1.5 & 3507\\ 
   \hline
    starling-lm:7b~\cite{starling2023} & 4 & 7 & 4.1 & 5183\\

    starling-lm:7b-alpha-q2\_K & 2 & 7 & 3.1 & 4285\\

    starling-lm:7b-alpha-q3\_K\_L & 3 & 3 & 3.8 & 4989\\
   \hline
    vicuna:7b~\cite{vicuna2023} & 4 & 7 & 3.8 & 5371\\

    vicuna:7b-q2\_K & 2 & 7 & 2.8 & 4411\\

    vicuna:7b-q2\_K\_L & 4 & 3 & 3.6 & 5115\\
    \hline
    zephyr:7b-beta~\cite{tunstall2023zephyr} & 4 & 7 & 4.1 & 5183\\

    zephyr:7b-beta-q2\_K & 2 & 7 & 3.1 & 4285\\

    zephyr:7b-beta-q3\_K\_L & 3 & 7 & 3.8 & 4989\\
    \Xhline{2\arrayrulewidth}\\
    \end{tabular}
    \caption{\scriptsize SLMs studied in this work.}
    \label{tab:slm_studied}
    \end{table}

\section{Evaluation} 
\label{sec:eval}
We leverage the SLaM methodology and toolset to investigate replacing OpenAI LLM with self-hosted SLMs for the production use case described in Section~\ref{sec:problem}.
In this section, we present the result of this case study and key observations and insights.
\subsection{Can SLMs generate responses as high-quality as OpenAI LLMs?}
\label{subsec:human-eval}

    \begin{figure}[t]
        \centering
        \includegraphics[width=\columnwidth]{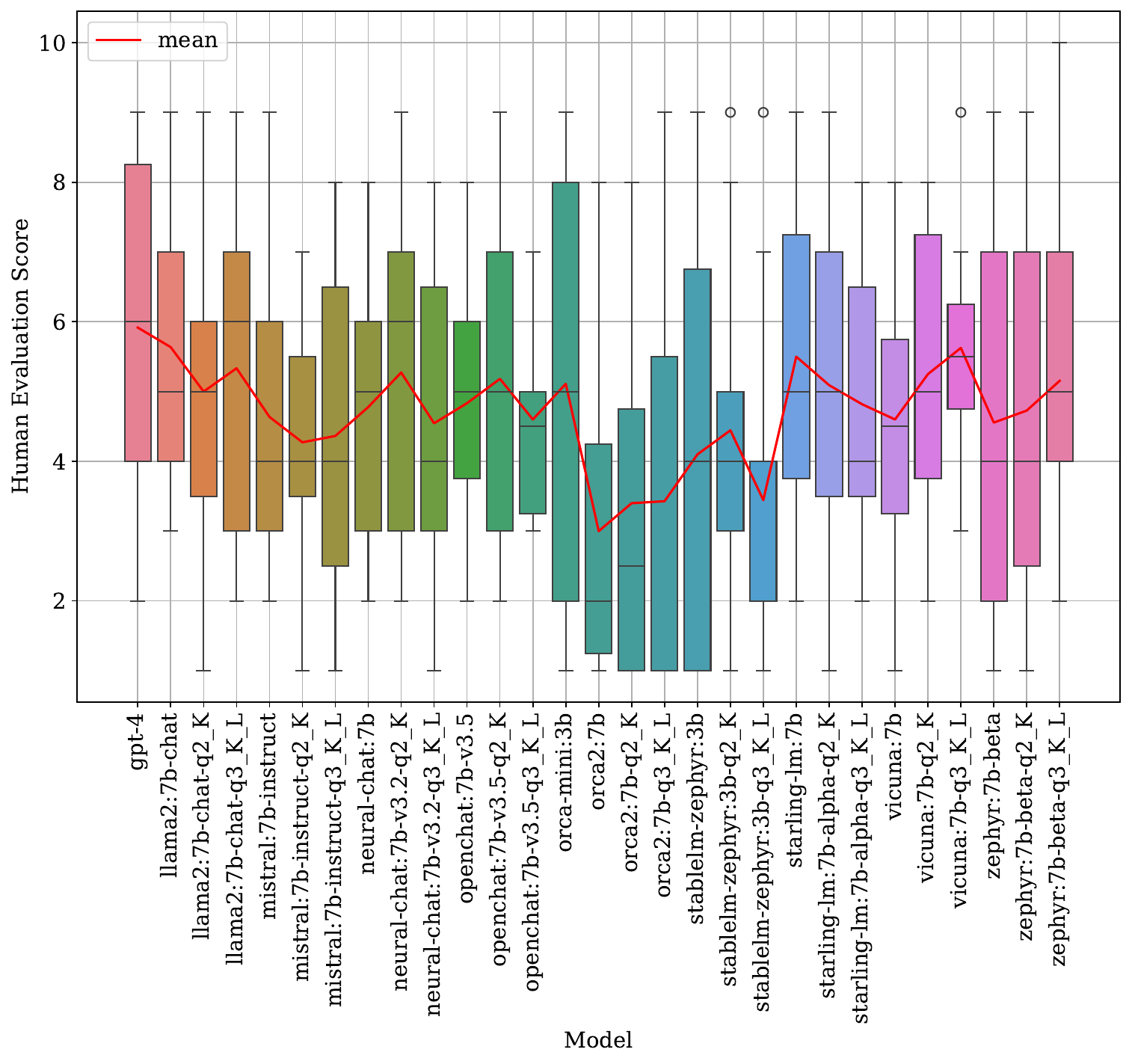}
        \caption {Response quality of GPT-4 and SLMs, as rated by human reviewers. Score distribution (boxes) and mean score (line) of each model are shown.}
        \label{fig:human_evaluation_boxplot}
    \end{figure} 

We first investigate the quality of the responses generated by the SLMs and compare them against OpenAI GPT-4.
Figure~\ref{fig:human_evaluation_boxplot} shows the scores of each model's responses rated by the team of human reviewers, following the evaluation process described in Section~\ref{subsec:human_eval_methodolodgy}.
The reviewers include product users and university students.
The boxes show the distribution of scores of each model, and the line represents the mean scores of each model across all human reviewers.
We observe that, while GPT-4 (1st box on the left) has the highest mean score (5.9), several SLMs achieved scores close to GPT-4, namely Vicuna:7b-q3 (5.6), Starling-lm:7b-q4 (5.1) and neural-chat:7b-q2 (5.3).
This demonstrates that certain SLMs can indeed generate responses with quality competitive to OpenAI LLMs.
In addition, we find that many quantized models perform similarly or even better than their base models.
This is particularly encouraging for applying SLMs in production because of the smaller size of quantized SLMs.

To demonstrate the response quality of SLMs, we showcase below sample responses of the daily pep talk feature from GPT-4 and two of the highest-rated SLMs, Starling-LM:7B and Vicua:7b-q3.

\vspace{2pt}
{
\scriptsize
\fontfamily{cmbr}\selectfont
\begin{spacing}{0.9}
\textbf{GPT-4 --} Good morning! You had a productive day last time, accomplishing both your daily rituals. Today, we're starting strong with planning the day, a task that aligns perfectly with your goal to get back into your daily routine. Focused tasks such as stress testing the openAI assistants API, preparing a demo for the RAG prototype, and improving the pep talk use cases, will help you dive deep into local LLMs and get the pep talk AI feature ready for productization. Don't forget your self-care tasks as well such as scheduling your orthopedic appointment, which addresses your goal to take care of your lower back issues. Let's make today a great one!

\vspace{1pt}

\textbf{starling-lm:7b --} Great job on completing the dailies on November 10th! Today, you have a full schedule ahead of you with key activities such as stress testing the OpenAI assistant API, preparing a demo for the RAG prototype, quantization and LORA of a model, and trying out GPT-4 Turbo. To align with your goals for this week, ensure that improving the pep talk use cases remains a priority and consider scheduling your orthopedic appointment sooner rather than later. Your daily rituals like planning, lifting, and medium scanning are crucial to maintaining good habits and staying focused. While some tasks may not be directly related to your goals, they contribute to overall progress and personal growth. Let's make today a productive one and keep moving forward with determination and enthusiasm!

\vspace{1pt}

\textbf{vicuna:7b-q3 --} Good morning! Based on your previous log-in and completed tasks, you were able to complete the daily meetings, which is fantastic. Today's focused tasks include planning for the day, stress testing the openAI assistants API, preparing a demo for the RAG prototype, trying quantization and LORA a model, trying out GPT-4-turbo and DALL-E 3, scheduling car service, improving the pep talk use cases, sending the consultation report, working on the AI project, and scheduling an orthopedic appointment. Your rituals for today include Daily, Zero Inbox, Review Elastic Logs, and lifting weights. These recurring tasks help you build good habits and maintain your workflow responsibilities. In addition, you set a few goals for yourself this week, which are to get the pep talk AI feature ready for productization, get back into your daily routine and workflow, take care of lower back issues, and dive deeper into local LLMs. Based on your focused tasks and rituals, it seems that you have a great start to your day ahead. You are taking care of some important projects and maintaining good habits. Good luck with achieving your goals this week!

\end{spacing}
}

In addition to the top-performing models, many of the studied SLMs achieve comparable performance and have scores close to the top models.
This suggests SLMs indeed can generate responses with quality close to OpenAI LLMs.
On the other hand, we do observe that a small subset of SLMs exhibit clearly inferior response quality compared to the rest of the group.
Specifically, the orca2:7b models (base and quantized) and stablelm-zephyr:3b-q3 scored significantly worse than the rest of the models.

\noindent\textbf{Key Observation:} Many SLMs, including their quantized versions, are able to generate responses of similar quality as OpenAI GPT-4, while a small set of SLMs are the clear low-performing options compared to the rest of the group.

\subsection{Can the evaluation process be automated?}
\label{subsec:gpt-scorer}
Human evaluation can be labor-intensive and time-consuming and are not always viable.
We explore the feasibility of leveraging GPT model as an evaluator and similarity metrics to automate the response quality scoring.
We study the usefulness of the GPT-based evaluators (~\ref{subsec:gpt-evaluators}) and semantic similarity approach (~\ref{subsec:similarity-score}).

    \begin{figure}[t!]
        \centering
        \includegraphics[width=\columnwidth]{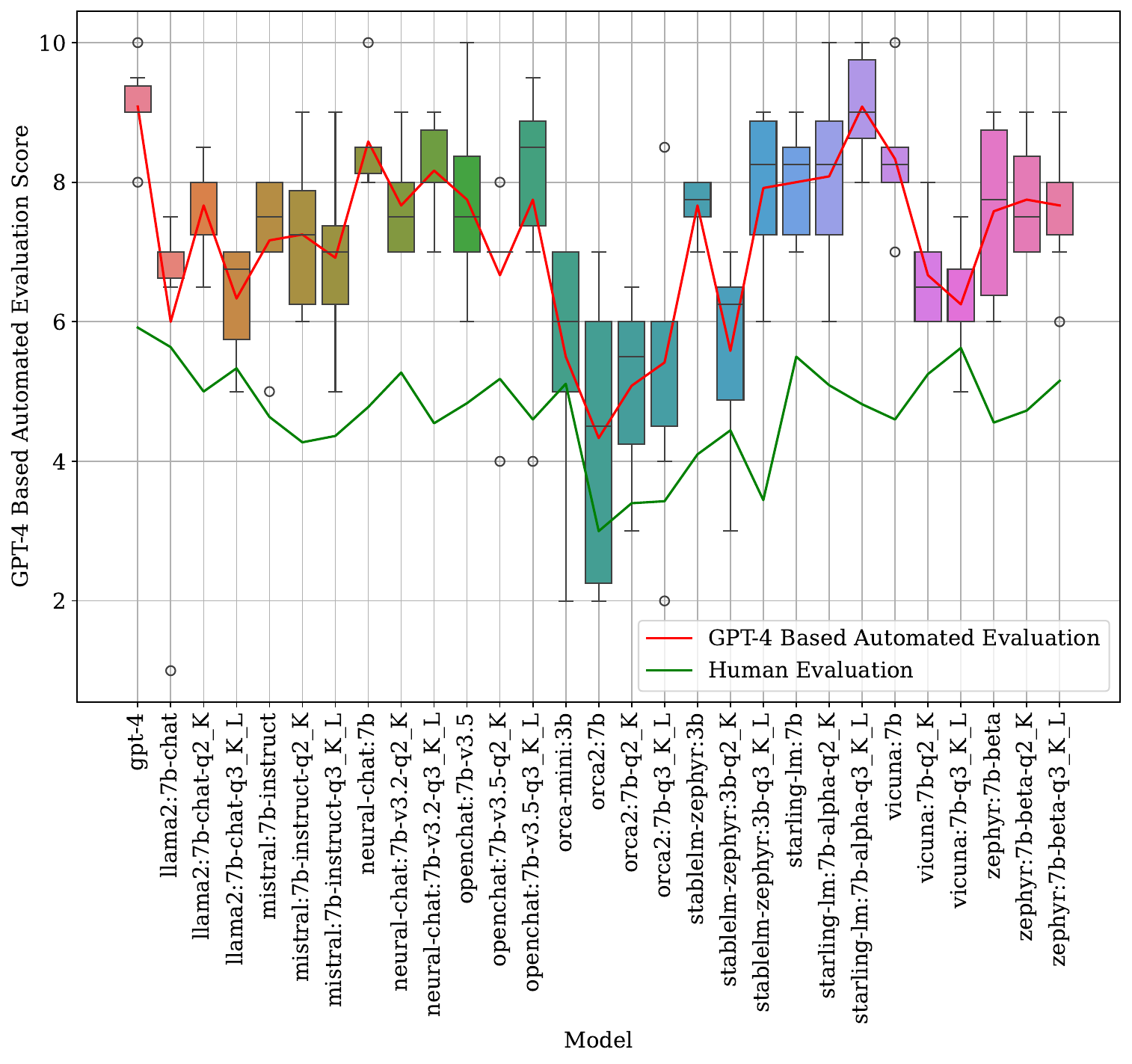}
        \caption{Distribution of scores given to 10 samples of responses from each model by the GPT-Scorer, compared with human evaluation scores.}
        \label{fig:gpt4_based_evaluation_scorer_boxplot}
    \end{figure}
        \begin{figure}[t!]
        \centering
        \includegraphics[width=0.8\columnwidth]{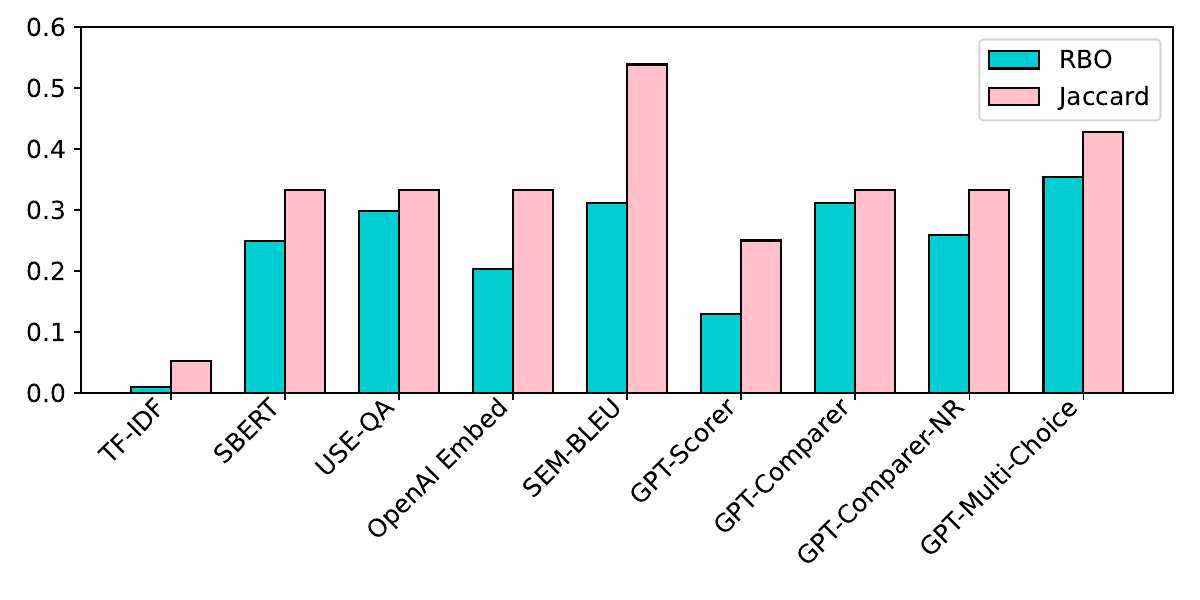}
        \caption {Jaccard index and Ranked-biased Overlap (RBO) of bottom 10 SLMs ranked by each automated evaluation method compared to human evaluation. Two variation of GPT-Comparers are tested here. GPT-Comparer explicitly prompts for reasoning while GPT-Comparer-NR does not.}
        \label{fig:jaccard-bot-10}
    \end{figure} 
    
\noindent\textbf{GPT-Scorer} Figure~\ref{fig:gpt4_based_evaluation_scorer_boxplot} shows the scores of response from each model as rated by a GPT-4 model with the instruction prompt shown in Section~\ref{subsec:gpt-evaluators}.
The boxes show the distribution of scores, and the two lines show the mean score of human evaluation and GPT-Scorer.
We observe that GPT-Scorer evaluation follows a similar trend as the human evaluation, indicating a similar relative model preference profile.
Furthermore, it is able to identify the same subset of inferior models as the human evaluation.
This suggests that prompting GPT-4 to score model response quality in the same style as human evaluation can provide model quality insights similar to human evaluation.

\noindent\textbf{Key Observation:} Scoring responses via prompting GPT-4 as an evaluator arrives at similar model quality preferences as human evaluation.


\begin{figure*}[t] 
  \begin{minipage}[b]{0.32\linewidth}
    \centering
    \includegraphics[width=\linewidth]{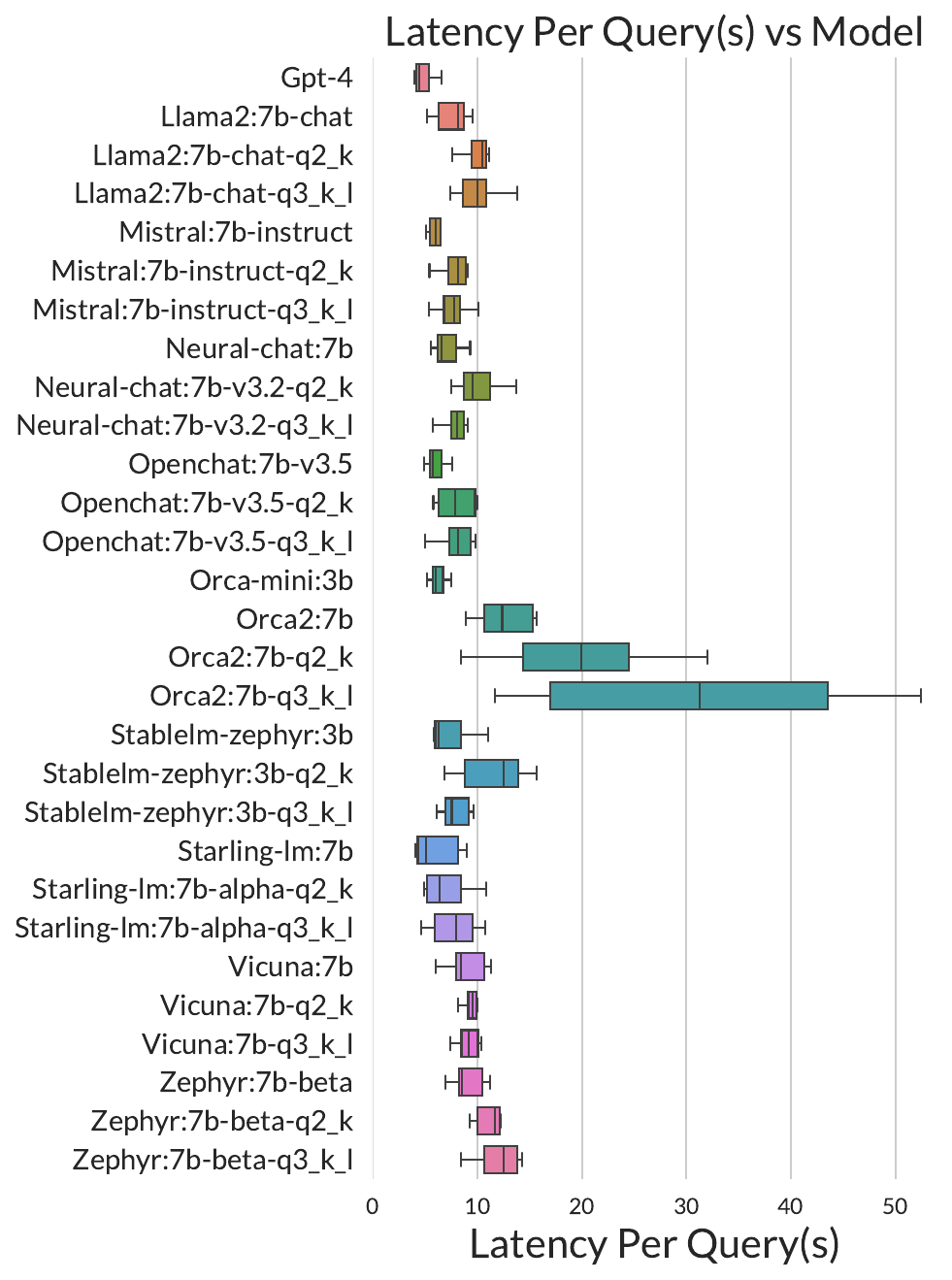}
    \caption{Latency per request of SLMs and OpenAI GPT-4 API. Distribution of 10 requests is shown here.}
    \label{fig:latency_per_request}
  \end{minipage}
  \hfill 
  \begin{minipage}[b]{0.32\linewidth}
    \centering
    \includegraphics[width=\linewidth]{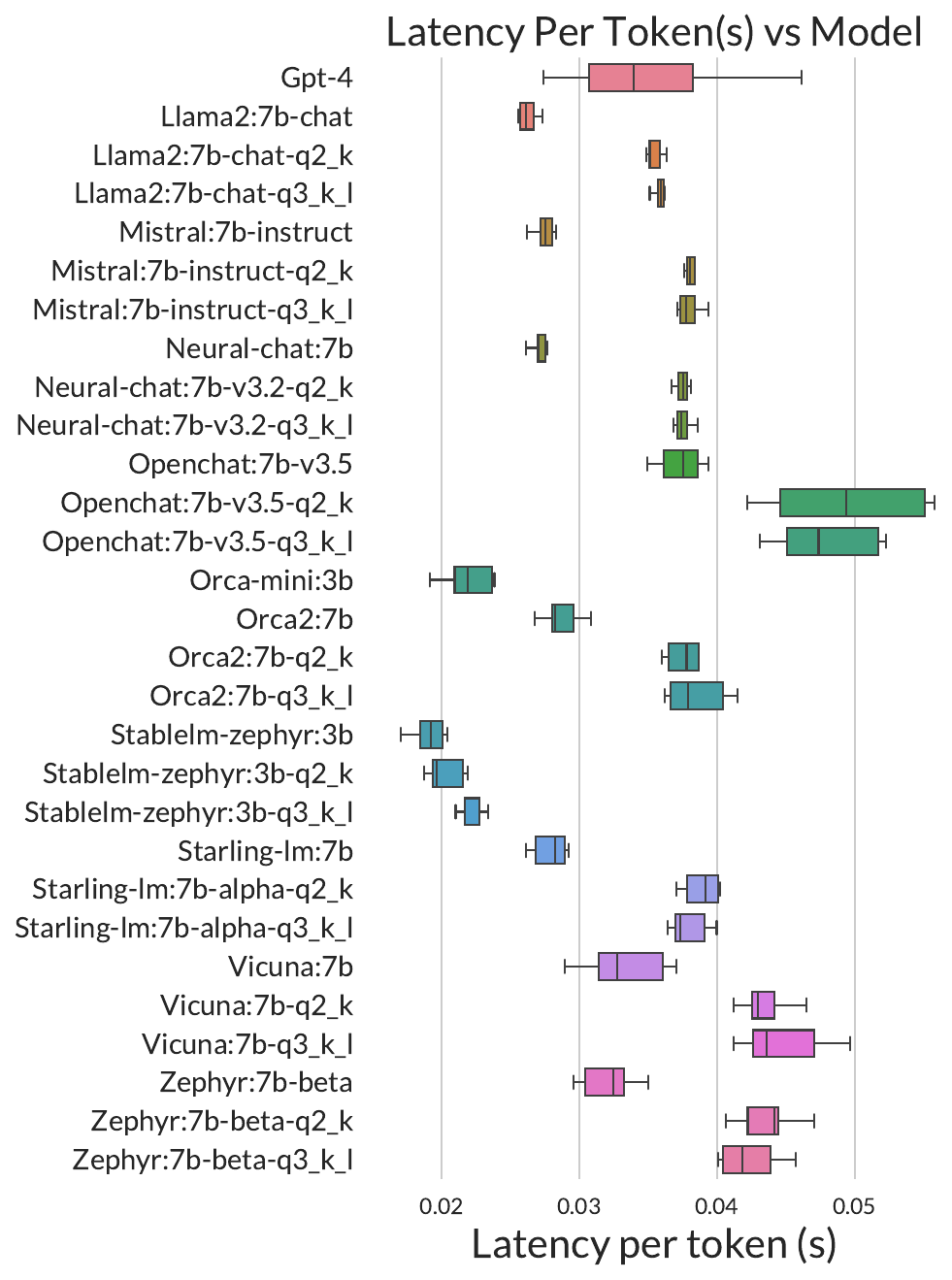}
    \caption{Latency per token of SLMs and OpenAI GPT-4 API. Distribution of 10 requests is shown here.}
    \label{fig:latency_per_token}
  \end{minipage}
  \hfill 
  \begin{minipage}[b]{0.32\linewidth}
    \centering
    \includegraphics[width=\linewidth]{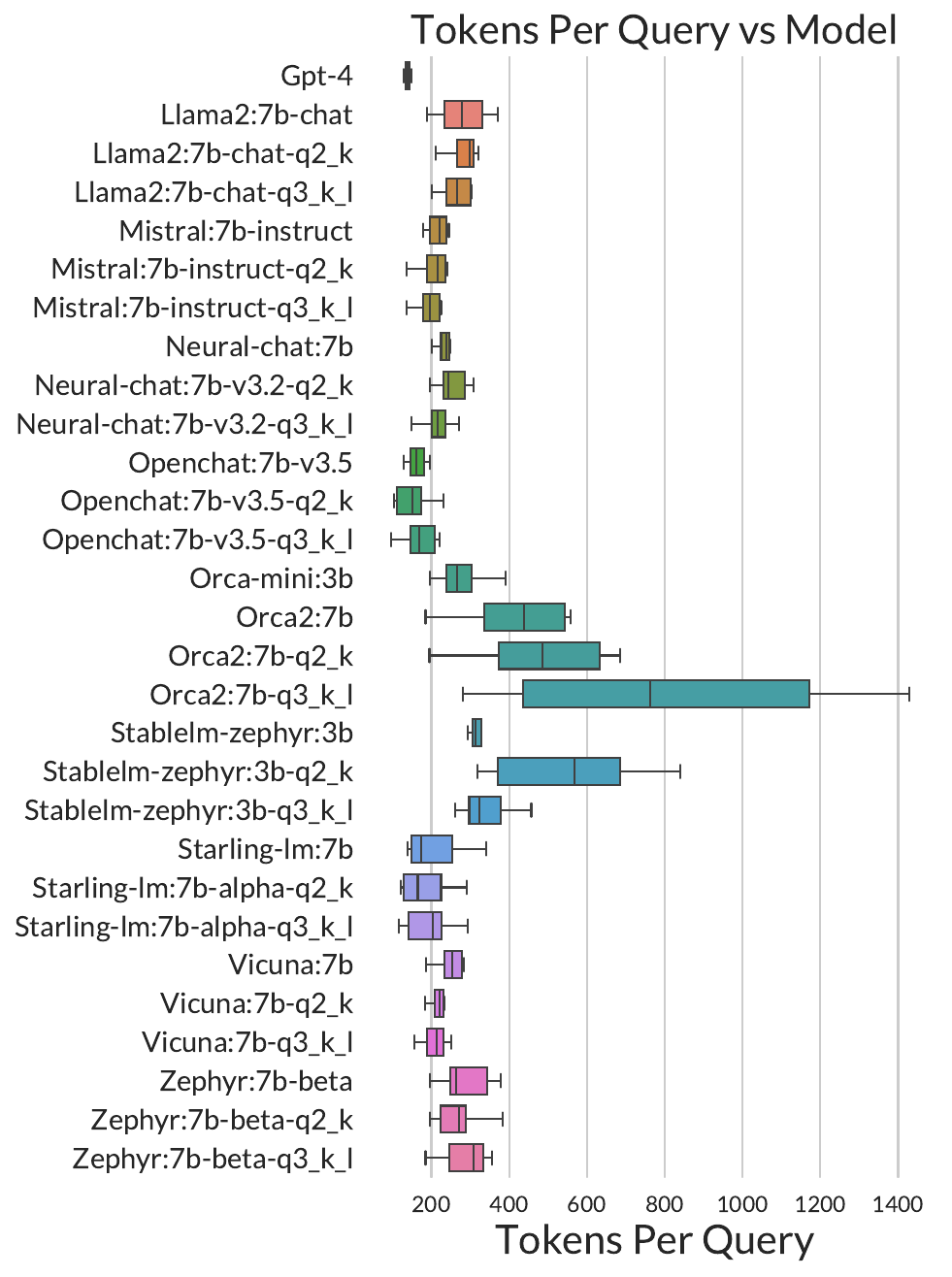}
    \caption{Number of tokens generated by SLMs and OpenAI GPT-4 API. Distribution of 10 requests with the same prompt shown here.}
    \label{fig:token_per_query}
  \end{minipage}
\end{figure*}

\noindent\textbf{GPT-based and Similarity-based Auto Evaluation} 
We now compare across the automated evaluators, including the GPT-based evaluators and similarity scoring method.
As discussed in Section~\ref{subsec:human-eval}, many SLMs provide competitive response quality to OpenAI while a small subset of SLMs appear clearly inferior.
Being able to quickly identify the low-performing models and remove them from consideration can greatly accelerate the SLM model evaluation and selection process by focusing expensive human effort on identifying the best models.
We focus on characterizing how well each auto-evaluation method can identify low-performing SLMs.
Figure~\ref{fig:jaccard-bot-10} shows the Jaccard index and Ranked-biased Overlap (RBO) with uniform weighting for each automated evaluation method (5 similarity scoring and 4 GPT-based) for the bottom 10 SLMs in their ranking compared to the human evaluation.
A higher Jaccard index and RBO for an automated evaluation method indicates that it ranks the model more similar to that of human rankings.
First, we observe that most methods have a relatively low RBO, indicating some disagreement in the order of the models ranked in the bottom 10 between human evaluation and the automated methods.
On the other hand, we observe that sem-bleu has the highest Jaccard index, meaning that its judgment of low-performing models is the most aligned with human ranking.
Sem-bleu, as described in Section~\ref{subsec:similarity-score}, combines sentence-level (SBERT) and token-level (BLEU) similarity, shares the same 7 out of 10 model choices with human evaluation in its bottom 10 rankings, albeit with different ordering as indicated by a low RBO score.

\noindent\textbf{Key Observation:} Automated evaluation methods can be effective in identifying low-performing models, accelerating the process of SLMs evaluation by focusing human evaluation on top models. Specifically, sem-bleu identifies low-performing models most similar to human evaluation.

\subsection{Are self-hosted SLMs faster than OpenAI APIs?}

\textbf{Per-request Latency --} Figure~\ref{fig:latency_per_request} shows the per-request latency of the SLMs and OpenAI GPT-4.
Distributions of 10 requests with the same input are shown here.
While GPT-4 (1st row) has the lowest per-request latency, most SLMs provide competitive latency performance.
Specifically, the mean request latency of mistral:7b-instruct, orca-mini:3b and stablelm-zephyr:3b are within $<$1 second of GPT-4's latency.

\textbf{Per-token Latency -} Language models generate its output one token at a time, and therefore, the end-to-end request latency depends on the number of tokens the model generates.
To further understand the performance characteristics of these SLMs, we characterize the number of tokens in the response generated by each model (Figure~\ref{fig:token_per_query}) and the per-token latency of each model (Figure~\ref{fig:latency_per_token}).
Figure~\ref{fig:token_per_query} shows that the Orca 2 models and StableLM-zephyr:3b-q2 have large variations in their response length.
This explains the wide distribution of per-request latency for these models observed in Figure~\ref{fig:latency_per_request}.
GPT-4's response length has a much lower variance than the SLMs, indicating its responses more consistently follow the prompt's instructions.
Furthermore, Figure~\ref{fig:latency_per_token} shows that many SLMs, such as StableLM-zephyr-3b, mistral-7b, and starlingLM, are faster at generating tokens than OpenAI GPT-4.

\noindent\textbf{Key Observation:} Self-hosted SLMs can achieve similar or better latency performance as OpenAI GPT-4 in per token generation. GPT-4 is more consistent in following response length control instructions in the prompt, providing more consistent performance on the request level, while some SLMs have a wider distribution of per-request latency.

\subsection{Are self-hosted SLMs more reliable than OpenAI APIs?}
\begin{figure}
    \centering
    \includegraphics[width=\columnwidth]{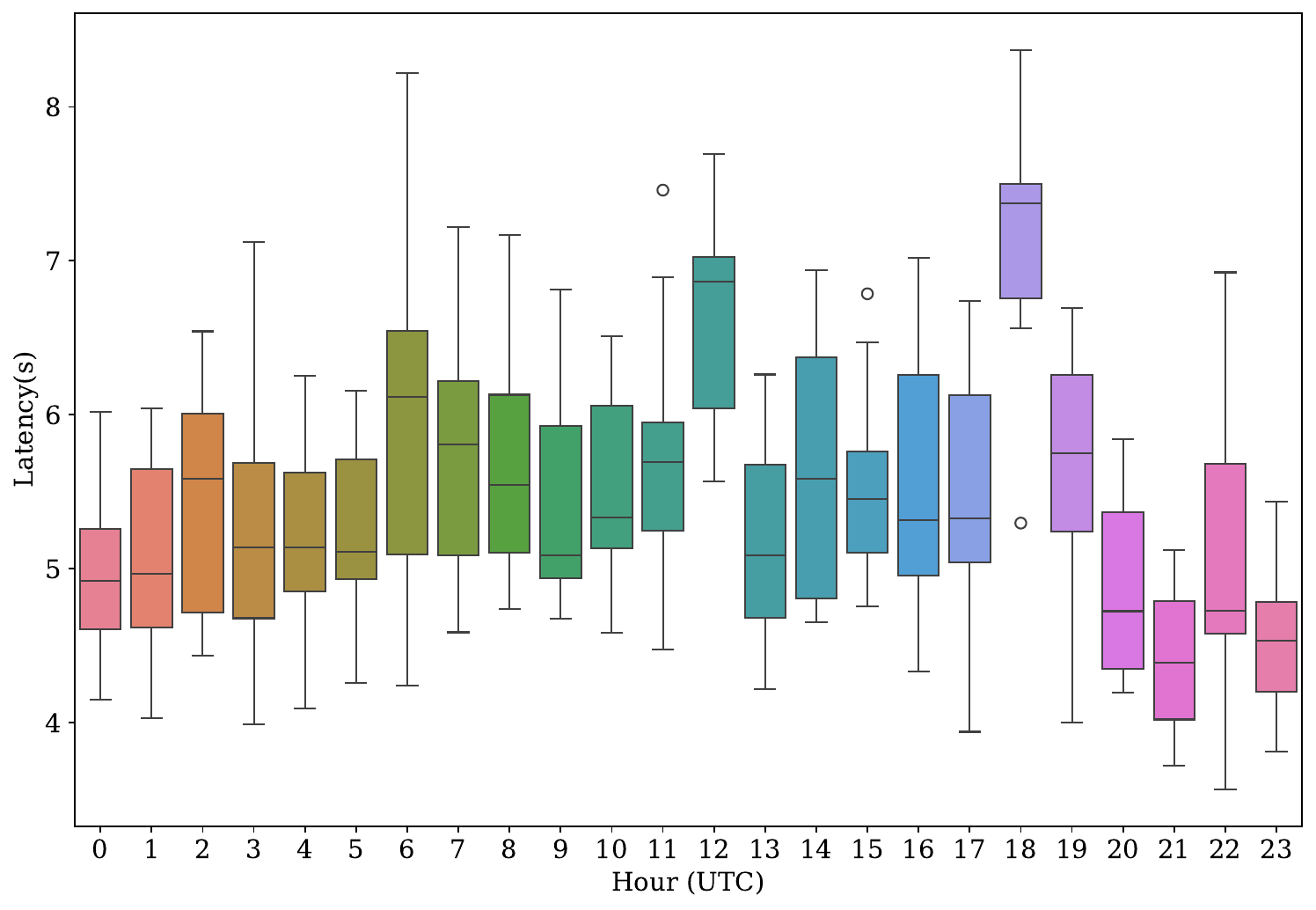}
    \caption {Latency distribution of OpenAI's API over the course of a 24-hour period.}
    \label{fig:openai-24hours}
\end{figure} 
\begin{figure}
    \centering
    \includegraphics[width=\columnwidth]{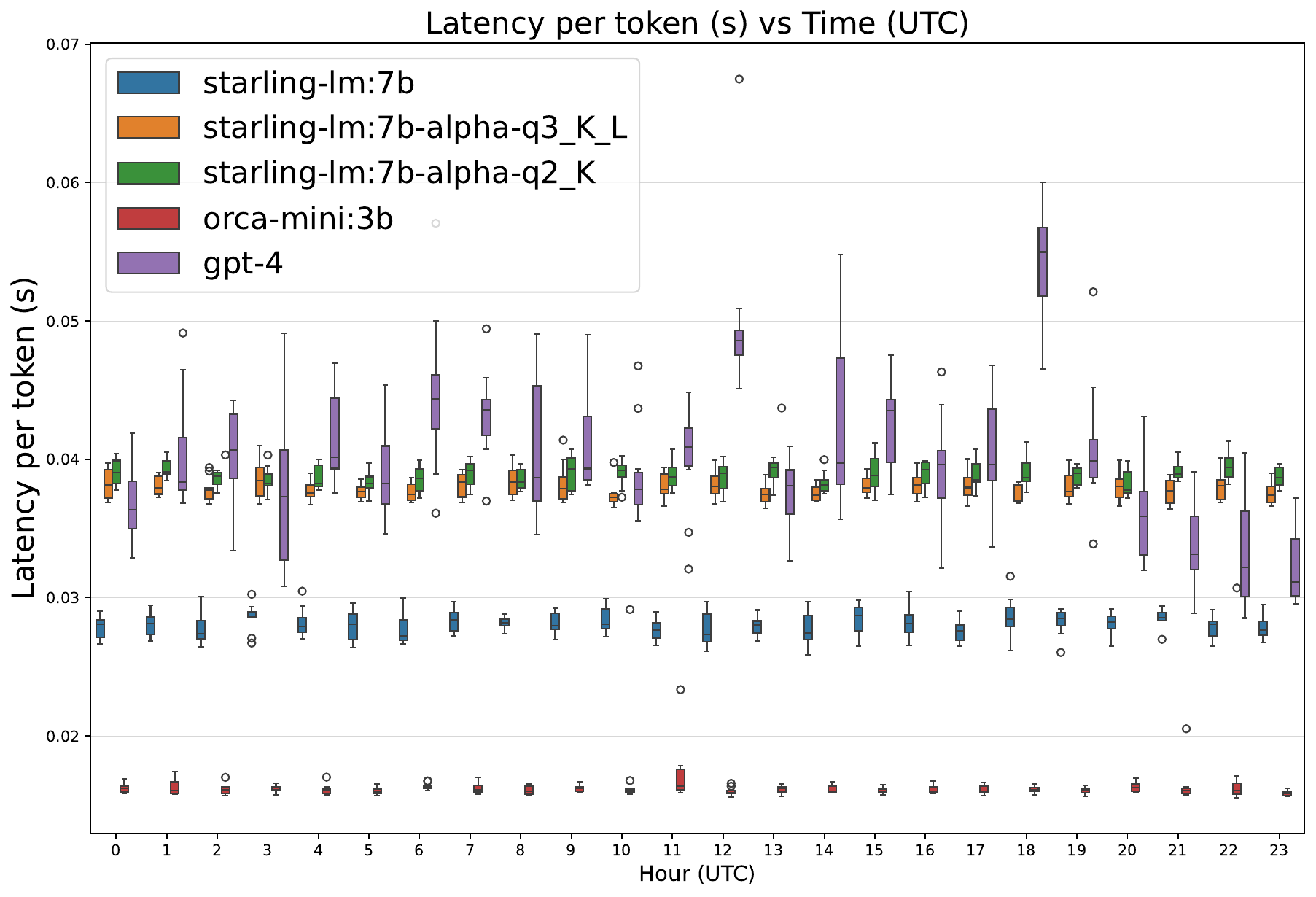}
    \caption {Latency distribution of OpenAI's GPT-4 API, Starling-lm:7b, and Orca-mini:3b over the course of a 24-Hour Period.}
    \label{fig:slms-24hours}
\end{figure} 

We characterize the variability in the performance of OpenAI APIs and self-hosting SLMs at different times of the day.
Figure~\ref{fig:openai-24hours} shows how the request latency of OpenAI API varies over the course of 24 hours.
Ten requests are measured at each hour, and their distributions are shown as boxes.
The same input and request parameters are used across all requests.
We observe significant variations in request latency for OpenAI at different times of the day, where per-request latency ranges from 3.4 seconds to 8.6 seconds.
This large variability presents a significant challenge in production settings, where predictability is crucial.

Figure~\ref{fig:slms-24hours} shows the latency variability over 24 hours of four selected SLMs hosted in AWS and compares them against OpenAI GPT-4.
StarlingLMs are selected because their response quality are among the best of the SLMs and Orca-mini-3b is selected because of its small size and yet strong response quality performance.
When measuring the SLMs latency, we warm up the machine with 10 requests first to ensure the model weights are memory-resident and we measure the latency of the subsequent 10 requests.
We observe that self-hosted SLMs have a significantly more consistent request latency over the course of the 24 hours and a narrower request-to-request distribution within the same hour.

\noindent\textbf{Key Observation:} Self-hosting SLMs can significantly reduce performance variability in production compared to OpenAI APIs.

\subsection{Are self-hosted SLMs more cost-effective than OpenAI APIs?}
\begin{figure}
        \centering
        \includegraphics[width=0.8\columnwidth]{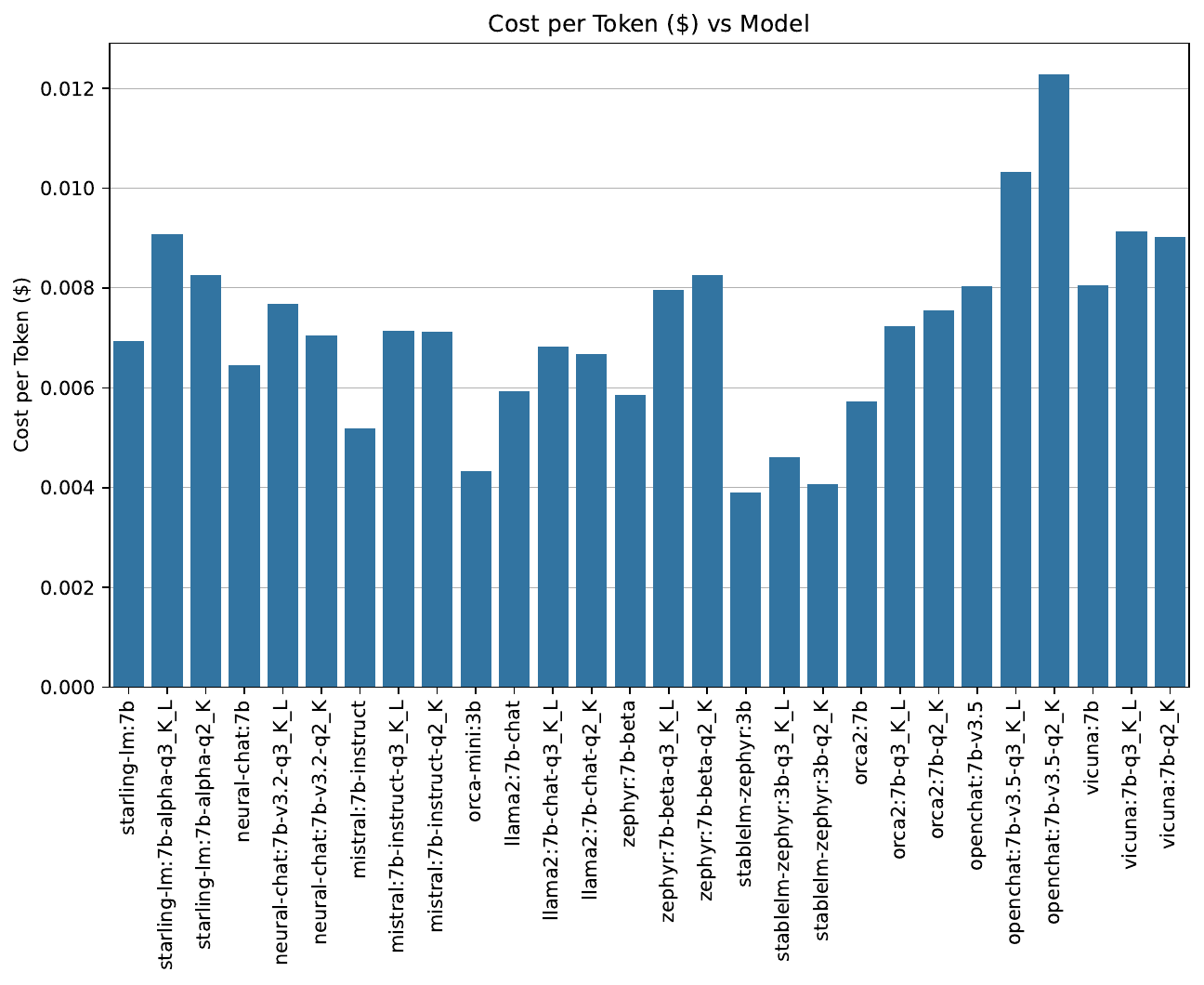}
        \caption {Estimated cost per 1K token generated for each SLM. GPT-4 incurs a cost of \$0.03 per input token and \$0.06 per completion token, whereas the self-hosted SLM models only incur costs associated with AWS. We assume 80\% utilization of the AWS node.}
        \label{fig:cost_per_1k_token}
    \end{figure}

\begin{figure}
        \centering
        \includegraphics[width=0.8\columnwidth]{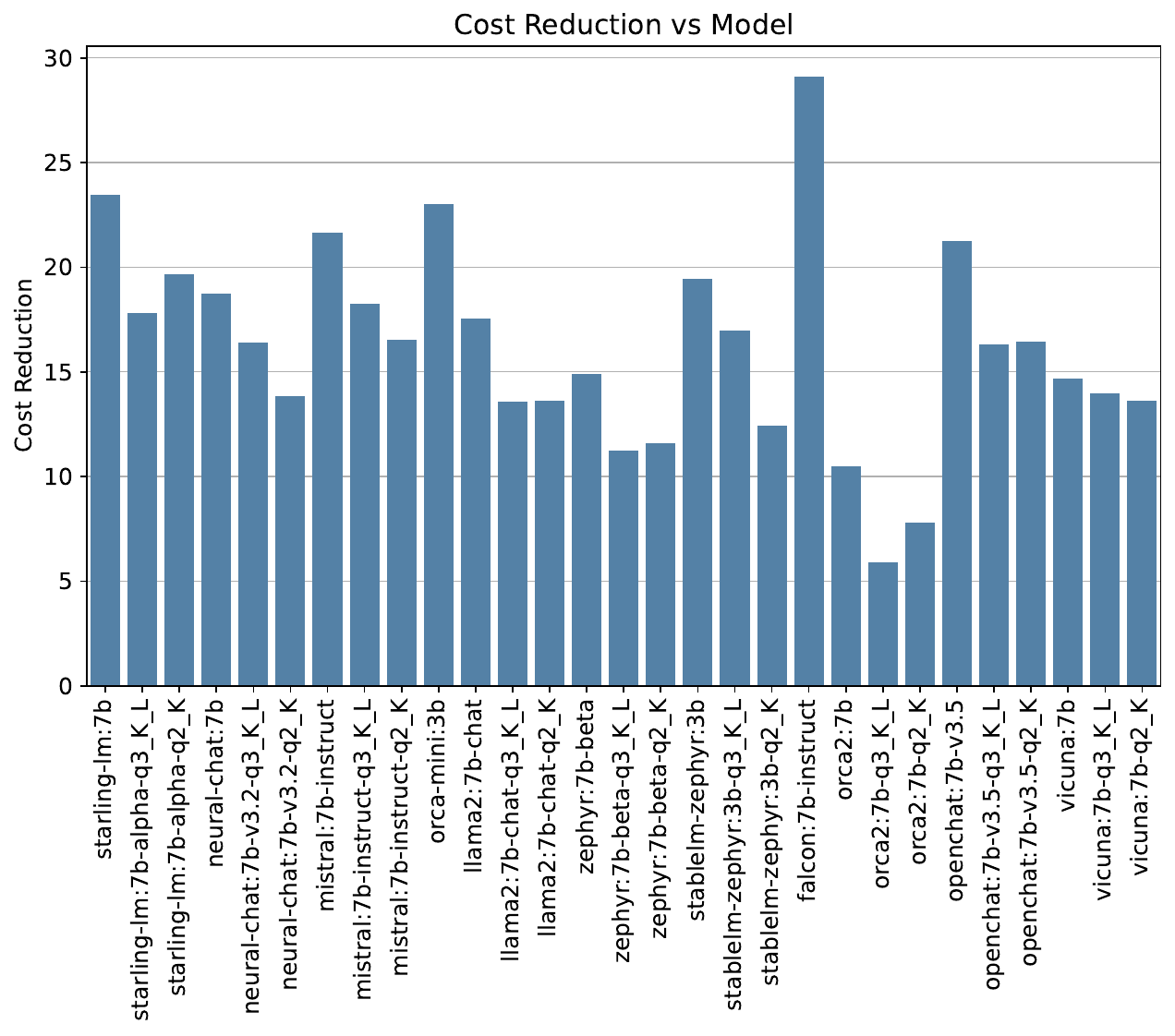}
        \caption {Cost reduction of each SLM (g4dn.xlarge AWS node at 80\% utilization) against OpenAI's GPT-4 API.}
        \label{fig:cost_reduction_per_query}
    \end{figure}

We compute the throughput of each SLM when running on an AWS instance at 80\% utilization. We then use the throughput to estimate the cost of self-hosting the SLMs. The per-1K token cost of self-hosting SLMs are shown in Figure~\ref{fig:cost_per_1k_token}. The corresponding OpenAI cost is \$0.09 (1K input and 1K output tokens). Figure~\ref{fig:cost_reduction_per_query} shows the cost reduction resulting from running the SLMs instead of querying OpenAI API. We picked 80\% utilization to be conservative in our comparison. The use of auto-scaling monitors utilization over time and automatically adds compute resources as needed, but its use also implies that an individual node will never be fully utilized. If the number of inferences is low, OpenAI APIs could have a cost advantage. On the other hand, the AWS instance could schedule containers unrelated to inference and amortize its cost in that case. Overall, we see a cost reduction of 5$\times$ to 29$\times$, depending on the model used.



%% file: reference.tex
\section{Related Works}

Zheng et al.~\cite{zheng2023judging} explore the usage and limitations of using LLMs as judges to evaluate models. They introduce a new benchmark focused on Question-Answering and a crowd rating platform for chatbots built using LLMs. In this work, we focus specifically on small language models and the SLaM methodology applies generally to language model tasks beyond QA and chatbots. We also go beyond response quality and present a holistic study of the readiness of SLMs to replace proprietary LLMs including latency and cost consideration.

We discuss prior works that address the challenges associated with large language models (LLMs) and ones that introduce innovative hardware and system solutions to accelerate them. 
Several work has investigated the performance of machine learning models focusing on the training aspects on various hardware designs~\cite{DBLP:conf/ispass/WonHRSSK23,DBLP:conf/ispass/MoolchandaniKRVEP23,10158216}. 
XRBench\cite{kwon2023xrbench} is a benchmarking suite for evaluating machine learning workloads related to extended reality. It assesses model dependency and concurrency for testing the real-time use of multi-model multi-task workloads. The concept of multi-DNN workloads is further developed by \cite{kwon2021heterogeneous} which introduces sub-accelerators to efficiently manage multiple subtasks concurrently.  
Xu et al. ~\cite{Xu2023SmallMA} discuss the challenges of tuning large language models (LLMs) like GPT-3 and GPT-4 due to their immense sizes and the inaccessibility of their weights.  Their proposed Super In-Context Learning (SuperICL) enables black-box LLMs to collaborate with locally finetuned smaller models. Liang, et. al\cite{liang2023unleashing}. demonstrate that LLMs can assist in developing quantum computing architectures on par with sophisticated quantum architecture search.

There are system design innovations that support and accelerate LLMs\cite{wu2023peta}\cite{blanaru2022enabling}\cite{song2023optimus}. For example, both FLAT ~\cite{kao2023flat} and SPRINT ~\cite{9923809} address the computational challenges of attention mechanisms in machine learning. FLAT introduces tailored dataflow optimization for attention mechanisms without altering their functionality. SPRINT is an accelerator that leverages the inherent parallelism of ReRAM crossbar arrays to compute attention scores approximately, and SUGAR is a Self-Adaptive Reconfigurable Array (SARA) with an integrated recommendation neural network (ADAPTNET) that enables runtime reconfiguration for optimized performance in accelerator architectures\cite{samajdar2021self}. Wang et al~\cite{10.1145/3567955.3567959} propose a novel technique to reduce data communication overheads in large deep learning models by overlapping communication with computation. This technique decomposes an identified original communication collective and the dependent computation operation into a sequence of fine-grained operations. Rouhani et al. ~\cite{10.1145/3579371.3589351} introduce Block Data Representations (BDR), a framework for exploring and evaluating a wide spectrum of narrow-precision formats for deep learning.  ~\cite{zhang2023hardware} presents a framework to evaluate hardware designs for accelerating LLM workloads.

There are also architectural innovations to accelerate ML and LLM models. VEGETA\cite{jeong2023vegeta} is an ISA extension for accelerating the computation of sparse matrix multiplication on the CPU for improving DNN workloads. 
Norm et al ~\cite{10.1145/3579371.3589350} discuss the design and implementation of Google’s domain-specific architecture (DSA) for machine learning models, including LLMs. They introduce SparseCores, dataflow processors that accelerate models that rely on embeddings. The performance, scalability, efficiency, and availability of TPU v4 make it an ideal vehicle for LLMs. Also, Seongmin et al ~\cite{9895626} presents a multi-FPGA acceleration appliance that executes GPT-2 model inference end-to-end with low latency and high throughput. Cong et al.~\cite{Guo_2023} proposes OliVe, an algorithm/architecture co-designed solution that adopts an outlier-victim pair (OVP) quantization that can be efficiently integrated into existing hardware accelerators using systolic arrays and tensor cores to achieve speedup and energy reduction with superior model accuracy.


%% file: main.bbl
\begin{thebibliography}{10}
\providecommand{\url}[1]{#1}
\csname url@samestyle\endcsname
\providecommand{\newblock}{\relax}
\providecommand{\bibinfo}[2]{#2}
\providecommand{\BIBentrySTDinterwordspacing}{\spaceskip=0pt\relax}
\providecommand{\BIBentryALTinterwordstretchfactor}{4}
\providecommand{\BIBentryALTinterwordspacing}{\spaceskip=\fontdimen2\font plus
\BIBentryALTinterwordstretchfactor\fontdimen3\font minus \fontdimen4\font\relax}
\providecommand{\BIBforeignlanguage}[2]{{%
\expandafter\ifx\csname l@#1\endcsname\relax
\typeout{** WARNING: IEEEtran.bst: No hyphenation pattern has been}%
\typeout{** loaded for the language `#1'. Using the pattern for}%
\typeout{** the default language instead.}%
\else
\language=\csname l@#1\endcsname
\fi
#2}}
\providecommand{\BIBdecl}{\relax}
\BIBdecl

\bibitem{openai_chatgpt}
``{OpenAI} blog: {ChatGPT},'' \url{https://openai.com/blog/chatgpt}, 2023.

\bibitem{dowling2023chatgpt}
M.~Dowling and B.~Lucey, ``Chatgpt for (finance) research: The bananarama conjecture,'' \emph{Finance Research Letters}, vol.~53, p. 103662, 2023.

\bibitem{openai_introducing_chatgpt_whisper_apis}
\BIBentryALTinterwordspacing
{OpenAI}. (Year Accessed) {Introducing ChatGPT and Whisper APIs}. [Online]. Available: \url{https://openai.com/blog/introducing-chatgpt-and-whisper-apis}
\BIBentrySTDinterwordspacing

\bibitem{theregister_chatgpt_outage_2023}
\BIBentryALTinterwordspacing
{The Register}. (2023) {Outage hits OpenAI's ChatGPT as servers stopped by 'Claude' glitch}. [Online]. Available: \url{https://www.theregister.com/2023/11/08/outage\_chatgpt\_openai\_claude/}
\BIBentrySTDinterwordspacing

\bibitem{touvron2023llama}
H.~Touvron, L.~Martin, K.~Stone, P.~Albert, A.~Almahairi, Y.~Babaei, N.~Bashlykov, S.~Batra, P.~Bhargava, S.~Bhosale \emph{et~al.}, ``Llama 2: Open foundation and fine-tuned chat models,'' \emph{arXiv preprint arXiv:2307.09288}, 2023.

\bibitem{jiang2023mistral}
A.~Q. Jiang, A.~Sablayrolles, A.~Mensch, C.~Bamford, D.~S. Chaplot, D.~de~las Casas, F.~Bressand, G.~Lengyel, G.~Lample, L.~Saulnier, L.~R. Lavaud, M.-A. Lachaux, P.~Stock, T.~L. Scao, T.~Lavril, T.~Wang, T.~Lacroix, and W.~E. Sayed, ``Mistral 7b,'' 2023.

\bibitem{neuralchat}
``neuralchat model listing,'' \url{https://huggingface.co/Intel/neural-chat-7b-v1-1}, 2023.

\bibitem{openllms23}
\BIBentryALTinterwordspacing
G.~Wang, S.~Cheng, Q.~Yu, and C.~Liu, ``{OpenLLMs: Less is More for Open-source Models},'' 7 2023. [Online]. Available: \url{https://github.com/imoneoi/openchat}
\BIBentrySTDinterwordspacing

\bibitem{mitra2023orca}
A.~Mitra, L.~D. Corro, S.~Mahajan, A.~Codas, C.~Simoes, S.~Agarwal, X.~Chen, A.~Razdaibiedina, E.~Jones, K.~Aggarwal, H.~Palangi, G.~Zheng, C.~Rosset, H.~Khanpour, and A.~Awadallah, ``Orca 2: Teaching small language models how to reason,'' 2023.

\bibitem{stablelm}
``Stabilitai: Stablelm,'' \url{https://github.com/Stability-AI/StableLM}, 2023.

\bibitem{starling2023}
B.~Zhu, E.~Frick, T.~Wu, H.~Zhu, and J.~Jiao, ``Starling-7b: Improving llm helpfulness and harmlessness with rlaif,'' November 2023.

\bibitem{vicuna2023}
\BIBentryALTinterwordspacing
W.-L. Chiang, Z.~Li, Z.~Lin, Y.~Sheng, Z.~Wu, H.~Zhang, L.~Zheng, S.~Zhuang, Y.~Zhuang, J.~E. Gonzalez, I.~Stoica, and E.~P. Xing, ``Vicuna: An open-source chatbot impressing gpt-4 with 90\%* chatgpt quality,'' March 2023. [Online]. Available: \url{https://lmsys.org/blog/2023-03-30-vicuna/}
\BIBentrySTDinterwordspacing

\bibitem{tunstall2023zephyr}
L.~Tunstall, E.~Beeching, N.~Lambert, N.~Rajani, K.~Rasul, Y.~Belkada, S.~Huang, L.~von Werra, C.~Fourrier, N.~Habib, N.~Sarrazin, O.~Sanseviero, A.~M. Rush, and T.~Wolf, ``Zephyr: Direct distillation of lm alignment,'' 2023.

\bibitem{shen2023efficient}
H.~Shen, H.~Chang, B.~Dong, Y.~Luo, and H.~Meng, ``Efficient llm inference on cpus,'' \emph{arXiv preprint arXiv:2311.00502}, 2023.

\bibitem{chang2023effective}
H.~Chang, H.~Shen, Y.~Cai, X.~Ye, Z.~Xu, W.~Cheng, K.~Lv, W.~Zhang, Y.~Lu, and H.~Guo, ``Effective quantization for diffusion models on cpus,'' \emph{arXiv preprint arXiv:2311.16133}, 2023.

\bibitem{wang2023openchat}
G.~Wang, S.~Cheng, X.~Zhan, X.~Li, S.~Song, and Y.~Liu, ``Openchat: Advancing open-source language models with mixed-quality data,'' \emph{arXiv preprint arXiv:2309.11235}, 2023.

\bibitem{mukherjee2023orca}
S.~Mukherjee, A.~Mitra, G.~Jawahar, S.~Agarwal, H.~Palangi, and A.~Awadallah, ``Orca: Progressive learning from complex explanation traces of gpt-4,'' \emph{arXiv preprint arXiv:2306.02707}, 2023.

\bibitem{mycaai}
\BIBentryALTinterwordspacing
``Myca.ai -- unleash your potential with ai-powered productivity.'' [Online]. Available: \url{https://www.myca.ai/}
\BIBentrySTDinterwordspacing

\bibitem{chang2023survey}
Y.~Chang and at~el., ``A survey on evaluation of large language models,'' 2023.

\bibitem{Ollivier2023}
M.~Ollivier and at~el., ``A deeper dive into chatgpt: history, use and future perspectives for orthopaedic research,'' \emph{Knee Surgery, Sports Traumatology, Arthroscopy: Official Journal of the ESSKA}, vol.~31, no.~4, pp. 1190--1192, 2023.

\bibitem{vaswani2023attention}
A.~Vaswani and at~el., ``Attention is all you need,'' 2023.

\bibitem{Radford2019LanguageMA}
\BIBentryALTinterwordspacing
A.~Radford and at~el., ``Language models are unsupervised multitask learners,'' 2019. [Online]. Available: \url{https://api.semanticscholar.org/CorpusID:160025533}
\BIBentrySTDinterwordspacing

\bibitem{shoeybi2020megatronlm}
M.~Shoeybi and at~el., ``Megatron-lm: Training multi-billion parameter language models using model parallelism,'' 2020.

\bibitem{OpenAI_GPT4_2023}
\BIBentryALTinterwordspacing
OpenAI, ``Gpt-4 technical report,'' \emph{ArXiv}, vol. abs/2303.08774, 2023. [Online]. Available: \url{https://arxiv.org/abs/2303.08774}
\BIBentrySTDinterwordspacing

\bibitem{openAIdevday}
\BIBentryALTinterwordspacing
``Join us for openai’s first developer conference,'' 2023. [Online]. Available: \url{https://openai.com/blog/announcing-openai-devday}
\BIBentrySTDinterwordspacing

\bibitem{liu2023llmqat}
Z.~L. et~al., ``Llm-qat: Data-free quantization aware training for large language models,'' 2023.

\bibitem{liu2023qllm}
J.~Liu, R.~Gong, X.~Wei, Z.~Dong, J.~Cai, and B.~Zhuang, ``Qllm: Accurate and efficient low-bitwidth quantization for large language models,'' 2023.

\bibitem{langchain-ai}
L.~AI, ``Langchain: Building applications with llms through composability,'' \url{https://github.com/langchain-ai/langchain}, 2023, accessed: Dec. 15, 2023.

\bibitem{llamaindex}
Llamaindex, ``Llamaindex, data framework for llm applications,'' \url{https://www.llamaindex.ai/}, 2024, accessed: April 13th, 2024.

\bibitem{prompt}
\BIBentryALTinterwordspacing
``Emnlp: Prompt engineering is the new feature engineering,'' 2022. [Online]. Available: \url{https://www.amazon.science/blog/emnlp-prompt-engineering-is-the-new-feature-engineering}
\BIBentrySTDinterwordspacing

\bibitem{10.1145/3560815}
\BIBentryALTinterwordspacing
P.~Liu, W.~Yuan, J.~Fu, Z.~Jiang, H.~Hayashi, and G.~Neubig, ``Pre-train, prompt, and predict: A systematic survey of prompting methods in natural language processing,'' \emph{ACM Comput. Surv.}, vol.~55, no.~9, jan 2023. [Online]. Available: \url{https://doi.org/10.1145/3560815}
\BIBentrySTDinterwordspacing

\bibitem{OpenAIforum}
\BIBentryALTinterwordspacing
``Openai: why are the api calls so slow? when will it be fixed,'' 2023. [Online]. Available: \url{https://community.openai.com/t/openai-why-are-the-api-calls-so-slow-when-will-it-be-fixed/148339/68?page=4}
\BIBentrySTDinterwordspacing

\bibitem{HuggingFace}
HuggingFace, ``{Hugging Face Models},'' \url{https://huggingface.co/models}, 2023, accessed: Dec. 15, 2023.

\bibitem{manning2009introduction}
C.~D. Manning, \emph{An introduction to information retrieval}.\hskip 1em plus 0.5em minus 0.4em\relax Cambridge university press, 2009.

\bibitem{DBLP:journals/corr/abs-1908-10084}
\BIBentryALTinterwordspacing
N.~Reimers and I.~Gurevych, ``Sentence-bert: Sentence embeddings using siamese bert-networks,'' \emph{CoRR}, vol. abs/1908.10084, 2019. [Online]. Available: \url{http://arxiv.org/abs/1908.10084}
\BIBentrySTDinterwordspacing

\bibitem{cer2018universal}
D.~Cer, Y.~Yang, S.-y. Kong, N.~Hua, N.~Limtiaco, R.~S. John, N.~Constant, M.~Guajardo-Cespedes, S.~Yuan, C.~Tar \emph{et~al.}, ``Universal sentence encoder,'' \emph{arXiv preprint arXiv:1803.11175}, 2018.

\bibitem{papineni2002bleu}
K.~Papineni, S.~Roukos, T.~Ward, and W.-J. Zhu, ``Bleu: a method for automatic evaluation of machine translation,'' in \emph{Proceedings of the 40th annual meeting of the Association for Computational Linguistics}, 2002, pp. 311--318.

\bibitem{ollam}
ollama, ``ollama,'' \url{https://github.com/ollama/ollama}, 2024, accessed: April 13th, 2024.

\bibitem{HuggingFaceH4}
HuggingFaceH4, ``{Open LLM Leaderboard},'' \url{https://huggingface.co/spaces/HuggingFaceH4/open\_llm\_leaderboard}, 2023, accessed: Dec. 15, 2023.

\bibitem{zheng2023judging}
L.~Zheng, W.-L. Chiang, Y.~Sheng, S.~Zhuang, Z.~Wu, Y.~Zhuang, Z.~Lin, Z.~Li, D.~Li, E.~P. Xing, H.~Zhang, J.~E. Gonzalez, and I.~Stoica, ``Judging llm-as-a-judge with mt-bench and chatbot arena,'' 2023.

\bibitem{DBLP:conf/ispass/WonHRSSK23}
\BIBentryALTinterwordspacing
W.~Won, T.~Heo, S.~Rashidi, S.~Sridharan, S.~Srinivasan, and T.~Krishna, ``Astra-sim2.0: Modeling hierarchical networks and disaggregated systems for large-model training at scale,'' in \emph{{IEEE} International Symposium on Performance Analysis of Systems and Software, {ISPASS} 2023, Raleigh, NC, USA, April 23-25, 2023}.\hskip 1em plus 0.5em minus 0.4em\relax {IEEE}, 2023, pp. 283--294. [Online]. Available: \url{https://doi.org/10.1109/ISPASS57527.2023.00035}
\BIBentrySTDinterwordspacing

\bibitem{DBLP:conf/ispass/MoolchandaniKRVEP23}
\BIBentryALTinterwordspacing
D.~Moolchandani, J.~Kundu, F.~Ruelens, P.~Vrancx, T.~Evenblij, and M.~Perumkunnil, ``Amped: An analytical model for performance in distributed training of transformers,'' in \emph{{IEEE} International Symposium on Performance Analysis of Systems and Software, {ISPASS} 2023, Raleigh, NC, USA, April 23-25, 2023}.\hskip 1em plus 0.5em minus 0.4em\relax {IEEE}, 2023, pp. 306--315. [Online]. Available: \url{https://doi.org/10.1109/ISPASS57527.2023.00037}
\BIBentrySTDinterwordspacing

\bibitem{10158216}
J.~Gómez-Luna, Y.~Guo, S.~Brocard, J.~Legriel, R.~Cimadomo, G.~F. Oliveira, G.~Singh, and O.~Mutlu, ``Evaluating machine learningworkloads on memory-centric computing systems,'' in \emph{2023 IEEE International Symposium on Performance Analysis of Systems and Software (ISPASS)}, 2023, pp. 35--49.

\bibitem{kwon2023xrbench}
H.~Kwon, K.~Nair, J.~Seo, J.~Yik, D.~Mohapatra, D.~Zhan, J.~Song, P.~Capak, P.~Zhang, P.~Vajda \emph{et~al.}, ``Xrbench: An extended reality (xr) machine learning benchmark suite for the metaverse,'' \emph{Proceedings of Machine Learning and Systems}, vol.~5, 2023.

\bibitem{kwon2021heterogeneous}
H.~Kwon, L.~Lai, M.~Pellauer, T.~Krishna, Y.-H. Chen, and V.~Chandra, ``Heterogeneous dataflow accelerators for multi-dnn workloads,'' in \emph{2021 IEEE International Symposium on High-Performance Computer Architecture (HPCA)}.\hskip 1em plus 0.5em minus 0.4em\relax IEEE, 2021, pp. 71--83.

\bibitem{Xu2023SmallMA}
C.~Xu and at~el., ``Small models are valuable plug-ins for large language models,'' \emph{ArXiv}, vol. abs/2305.08848, 2023.

\bibitem{liang2023unleashing}
Z.~Liang, J.~Cheng, R.~Yang, H.~Ren, Z.~Song, D.~Wu, X.~Qian, T.~Li, and Y.~Shi, ``Unleashing the potential of llms for quantum computing: A study in quantum architecture design,'' \emph{arXiv preprint arXiv:2307.08191}, 2023.

\bibitem{wu2023peta}
Z.~Wu, L.~Y. Dai, A.~Novick, M.~Glick, Z.~Zhu, S.~Rumley, G.~Michelogiannakis, J.~Shalf, and K.~Bergman, ``Peta-scale embedded photonics architecture for distributed deep learning applications,'' \emph{Journal of Lightwave Technology}, 2023.

\bibitem{blanaru2022enabling}
F.~Blanaru, A.~Stratikopoulos, J.~Fumero, and C.~Kotselidis, ``Enabling pipeline parallelism in heterogeneous managed runtime environments via batch processing,'' in \emph{Proceedings of the 18th ACM SIGPLAN/SIGOPS International Conference on Virtual Execution Environments}, 2022, pp. 58--71.

\bibitem{song2023optimus}
J.~Song, J.~Yim, J.~Jung, H.~Jang, H.-J. Kim, Y.~Kim, and J.~Lee, ``Optimus-cc: Efficient large nlp model training with 3d parallelism aware communication compression,'' in \emph{Proceedings of the 28th ACM International Conference on Architectural Support for Programming Languages and Operating Systems, Volume 2}, 2023, pp. 560--573.

\bibitem{kao2023flat}
S.-C. Kao, S.~Subramanian, G.~Agrawal, A.~Yazdanbakhsh, and T.~Krishna, ``Flat: An optimized dataflow for mitigating attention bottlenecks,'' in \emph{Proceedings of the 28th ACM International Conference on Architectural Support for Programming Languages and Operating Systems, Volume 2}, 2023, pp. 295--310.

\bibitem{9923809}
\BIBentryALTinterwordspacing
A.~Yazdanbakhsh and et~al., ``Sparse attention acceleration with synergistic in-memory pruning and on-chip recomputation,'' in \emph{2022 55th IEEE/ACM International Symposium on Microarchitecture (MICRO)}.\hskip 1em plus 0.5em minus 0.4em\relax Los Alamitos, CA, USA: IEEE Computer Society, oct 2022, pp. 744--762. [Online]. Available: \url{https://doi.ieeecomputersociety.org/10.1109/MICRO56248.2022.00059}
\BIBentrySTDinterwordspacing

\bibitem{samajdar2021self}
A.~Samajdar, M.~Pellauer, and T.~Krishna, ``Self-adaptive reconfigurable arrays (sara): Using ml to assist scaling gemm acceleration,'' \emph{arXiv preprint arXiv:2101.04799}, 2021.

\bibitem{10.1145/3567955.3567959}
S.~Wang and at~el., ``Overlap communication with dependent computation via decomposition in large deep learning models,'' in \emph{Proceedings of the 28th ACM International Conference on Architectural Support for Programming Languages and Operating Systems, Volume 1}, ser. ASPLOS 2023.\hskip 1em plus 0.5em minus 0.4em\relax Association for Computing Machinery, 2022, p. 93–106.

\bibitem{10.1145/3579371.3589351}
\BIBentryALTinterwordspacing
D.~Rouhani and et~al., ``With shared microexponents, a little shifting goes a long way,'' in \emph{Proceedings of the 50th Annual International Symposium on Computer Architecture}, ser. ISCA '23.\hskip 1em plus 0.5em minus 0.4em\relax New York, NY, USA: Association for Computing Machinery, 2023. [Online]. Available: \url{https://doi.org/10.1145/3579371.3589351}
\BIBentrySTDinterwordspacing

\bibitem{zhang2023hardware}
H.~Zhang, A.~Ning, R.~Prabhakar, and D.~Wentzlaff, ``A hardware evaluation framework for large language model inference,'' 2023.

\bibitem{jeong2023vegeta}
G.~Jeong, S.~Damani, A.~R. Bambhaniya, E.~Qin, C.~J. Hughes, S.~Subramoney, H.~Kim, and T.~Krishna, ``Vegeta: Vertically-integrated extensions for sparse/dense gemm tile acceleration on cpus,'' in \emph{2023 IEEE International Symposium on High-Performance Computer Architecture (HPCA)}.\hskip 1em plus 0.5em minus 0.4em\relax IEEE, 2023, pp. 259--272.

\bibitem{10.1145/3579371.3589350}
N.~Jouppi and et~al., ``Tpu v4: An optically reconfigurable supercomputer for machine learning with hardware support for embeddings,'' in \emph{Proceedings of the 50th Annual International Symposium on Computer Architecture}, ser. ISCA '23.\hskip 1em plus 0.5em minus 0.4em\relax Association for Computing Machinery, 2023.

\bibitem{9895626}
S.~Hong and at~el., ``Dfx: A low-latency multi-fpga appliance for accelerating transformer-based text generation,'' in \emph{2022 IEEE Hot Chips 34 Symposium (HCS)}, 2022, pp. 1--17.

\bibitem{Guo_2023}
\BIBentryALTinterwordspacing
C.~Guo and et~al., ``Olive: Accelerating large language models via hardware-friendly outlier-victim pair quantization,'' in \emph{Proceedings of the 50th Annual International Symposium on Computer Architecture}, ser. ISCA ’23.\hskip 1em plus 0.5em minus 0.4em\relax ACM, Jun. 2023. [Online]. Available: \url{http://dx.doi.org/10.1145/3579371.3589038}
\BIBentrySTDinterwordspacing

\end{thebibliography}
